\documentclass[oldversion]{aa} 
\usepackage{graphicx}
\usepackage{longtable}
\usepackage{txfonts}
\usepackage{rotating}
\usepackage{lscape}
\newcommand{\gsim}{\;\lower.6ex\hbox{$\sim$}\kern-7.75pt\raise.65ex\hbox{$>$}\;}
\newcommand{\lsim}{\;\lower.6ex\hbox{$\sim$}\kern-7.75pt\raise.65ex\hbox{$<$}\;}
\newcommand{\teff}{$T_{\rm eff}$}

\begin{document}
\title{The Na-O anticorrelation in horizontal branch stars. IV. M~22
\thanks{Based on observations collected at 
ESO telescopes under programmes 087.D-0230 and 091.D-0151}
\fnmsep\thanks{
   Tables 3, 4, 5, and 6 are only available in electronic form at the CDS via anonymous
   ftp to {\tt cdsarc.u-strasbg.fr} (130.79.128.5) or via
   {\tt http://cdsweb.u-strasbg.fr/cgi-bin/qcat?J/A+A/???/???}}
}

\author{
R.G. Gratton\inst{1},
S. Lucatello\inst{1},
A. Sollima\inst{2},
E. Carretta\inst{2},
A. Bragaglia\inst{2},
Y. Momany\inst{1,3},
V. D'Orazi\inst{4,5},
S. Cassisi\inst{6,7},
\and
M. Salaris\inst{8}}

\authorrunning{R.G. Gratton}
\titlerunning{Na-O in HB stars of M~22}

\offprints{R.G. Gratton, raffaele.gratton@oapd.inaf.it}

\institute{
INAF-Osservatorio Astronomico di Padova, Vicolo dell'Osservatorio 5, I-35122
 Padova, Italy
\and
INAF-Osservatorio Astronomico di Bologna, Via Ranzani 1, I-40127, Bologna, Italy
\and
European Southern Observatory, Alonso de Cordova 3107, Vitacura, Santiago, Chile 
\and
Department of Physics \& Astronomy, Macquarie University, Balaclava Rd., North Ryde, Sydney, NSW 2109, Australia
\and
Monash Centre for Astrophysics, School of Mathematical Sciences, Building 28, Monash University, VIC 3800, Australia
\and
INAF-Osservatorio Astronomico di Teramo, Via Collurania, Teramo, Italy
\and
Instituto de Astrofisica de Canarias, La Laguna, Tenerife, Spain
\and
Astrophysics Research Institute, Liverpool John Moores University, Twelve Quays House, Birkenhead, UK}

\date{}
\abstract{
We obtained high-resolution spectra for 94 candidate stars belonging to the 
HB of M~22 with FLAMES. Previous works have indicated that this cluster has 
split subgiant (SGB) and red giant branches (RGB) and hosts two different 
stellar populations, differing in overall metal abundance and both exhibiting 
a Na-O anti-correlation. The HB stars we observed span a restricted temperature 
range (7,800$<T_{\rm eff}<$11,000~K), where about 60\% of the HB stars of M~22 
are. Within our sample, we can distinguish three groups of stars segregated 
(though contiguous) in colours: Group 1 (49 stars) is metal-poor, N-normal, 
Na-poor and O-rich: our abundances for this (cooler) group match very well 
those determined for the primordial group of RGB stars (a third of the total) 
from previous studies. Group 2 (23 stars) is still metal-poor, but it is N- 
and Na-rich, though only very mildly depleted in O. We can identify this 
intermediate group as the progeny of the metal-poor RGB stars that occupy an 
intermediate location along the Na-O anti-correlation and include about 10\% 
of the RGB stars. The third group (20 stars) is metal-rich, Na-rich, and O-rich. 
This hotter group likely corresponds to the most O-rich component of the 
previously found metal-rich RGB population (a quarter of the total). We did 
not observe any severely O-depleted stars and we think that the progeny of these 
stars falls on the hotter part of the HB. Furthermore, we found that the 
metal-rich population is also over-abundant in Sr, in agreement with results for 
corresponding RGB and SGB stars. However, we do not find any significant 
variation in the ratio between the sum of N and O abundances to Fe. We do not 
have C abundances for our stars. There is some evidence of an enhancement of He 
content for Groups 2 and 3 stars ($Y=0.338\pm 0.014\pm 0.05$); the error bar due 
to systematics is large, but a consistent analysis of data for several GCs 
confirms that stars in these groups within M~22 are likely overabundant in He. 
We conclude that on the whole, our results agree with the proposition that 
chemical composition drives the location of stars along the HB of a GC. 
Furthermore, we found a number of fast rotators. They are concentrated in a 
restricted temperature range along the HB of M~22. Fast rotating stars might be 
slightly less massive and bluer than slowly rotating ones, but other
interpretations are possible.}
\keywords{Stars: abundances -- Stars: evolution --
Stars: Population II -- Galaxy: globular clusters -- Galaxy: individual: M22 }

\maketitle

\section{Introduction}

Low-mass core He-burning stars show a wide distribution in the colour-magnitude diagram 
of globular clusters (GCs) along the so-called horizontal branch (HB). This
distribution primarily reflects variations in masses and the chemical composition of
stars, with a minor but not negligible part being played by the stars'
evolution off their initial location on the zero age horizontal branch (ZAHB).
Full understanding of the reasons individual stars occupy a given position
along the ZAHB has still not been achieved, probably because several different
mechanisms are involved simultaneously. This constitutes the ``second
parameter problem" (Sandage \& Wildey 1967; van den Bergh 1967), and the first
parameter is  metallicity, which is responsible for most of the observational
variance (Sandage \& Wallerstein 1960; Faulkner 1966). Important progress
has been made thanks to the understanding that significant star-to-star
variations can be expected in the helium content within individual GCs, which
are made of different stellar  populations (see Ventura et al. 2001; Bedin et
al. 2004; D'Antona et al.  2005; Norris 2004; Piotto et al. 2005, following much
earlier  suggestions by e.g. Rood 1973 and Norris et al. 1981). Older and/or
He-richer stars are expected to leave the main sequence phase with lower
masses. If they lose a similar amount of mass as younger and/or He-normal stars
along the RGB, they are expected to be less massive, that is, bluer, when on
the HB. There is a broad correlation between extension of the HB and properties
very likely related to He abundances, such as the extension of the Na-O
anti-correlation (Na-rich and O-poor stars are expected to be more He-rich
than Na-poor and O-rich ones; Carretta et al. 2007), both driven mainly by
the total  mass of the GCs (Recio-Blanco et al. 2006). 

A few years ago, we (Gratton et  al. 2010; see also Dotter 2013 for a review of
this and later  contributions) presented a quite extensive re-analysis of the
distribution of stars along the HB of several tens GCs and found that a
combination of  variations in age (from cluster-to-cluster) and He-content (from
star-to-star  within a cluster), added to metallicity, may indeed explain most
of the variance in the HB morphology. This was achieved by adopting a simple
universal relation between the total mass lost along the RGB and metallicity,
with a small (but not negligible) star-to-star random contribution (one-two
hundredths of solar masses, about 10\% of the total mass lost along the RGB).
The presence of  this last random term most likely implies that even if the scenario
considered by Gratton et al. were broadly correct, something still needs to be
added in order to achieve very accurate predictions. Potential candidates,
whose importance has not yet been well established, include variations in the CNO/Fe
abundance ratio, core rotation, and binarity. The list of potential parameters
is even longer, see Fusi Pecci \& Bellazzini (1997) and Catelan (2009) for more
comprehensive  summaries. 

A corollary of the star-to-star He abundance variations explanation for the 
distribution of stars along the HB of an individual GC is that there should be
correlations between temperatures and chemical abundances, only partly fuzzed
by evolution off the ZAHB. Such correlations can be retrieved through
spectroscopy of HB stars. However, spectra of HB stars are difficult to
analyse. Ever since the pioneering work of Peterson (1983) we know that rotation can
be present on the BHB. Furthermore, abundances for stars hotter than about
11,000 K (the so-called  Grundahl-jump, Grundahl et al. 1999) are heavily
affected by diffusion and  radiative levitation (e.g. Behr et al. 1999, Mohler 
2001; Moni Bidin et al. 2006).

Villanova et al. (2009) first tried to connect spectroscopic determinations of
the composition of stars along the HB of NGC~6752 with the multiple population
scenarios and were also able to obtain information about He, though with 
non-negligible error bars, finding a low He abundance consistent with the
cosmological value, as expected for the kind of stars observed. In fact, He-rich
stars are  expected to be hotter than the Grundahl-jump in most old and
metal-poor GCs in order to avoid too bright HB's at the RR Lyrae colours, as
has been known for several  decades (see e.g. Iben 1968; Cassisi et al. 2003; Salaris
et al. 2004; and the  review by Catelan 2009). Two other papers on M~4 (Marino
et al. 2011a; Villanova et  al. 2012) have confirmed what has been found by Villanova et al.
(2009): red HB stars (RHB: that is, stars redder than the RR Lyrae  instability
strip) are Na-poor and O-rich, while BHB stars are Na-rich and O-poor. 
Furthermore, the observed BHB stars of M~4, which are amongst the warmest in that
cluster, are more He-rich than the stars observed in NGC~6752.

In this series of papers we present the analysis of wide samples of HB
stars  for a few important GCs. Table~\ref{t:tabsum} gives a summary of main
results obtained in published papers. In Gratton et al. (2011) we considered 
NGC~2808; as in M~4, which has similar metallicity, RHB stars  are O-rich,
but they show a spread in Na abundances correlated with temperature. Blue HB
stars cooler than the Grundahl-jump are (moderately) O-poor and Na-rich. These
results have been confirmed by Marino et al. (2013b) using higher S/N spectra,
from which they also derived He abundances and finding quite a high value of $Y\sim
0.34\pm 0.01\pm 0.05$ for BHB stars. Even He-richer stars should be present, but
they should be hotter than the Grundahl-jump (D'Antona et al. 2005), so
this cannot be verified directly. Gratton et al. (2012a) studied NGC~1851,
which is a complex GC with a split SGB (Milone et al. 2009), two populations slightly
differing in their Fe-content (Carretta et al. 2010, 2011), and related to the
bright (b-)SGB (metal-poor) and the faint (f-)SGB (metal-rich:  Gratton et al.
2012b), and two distinct Na-O anti-correlations (Carretta et al. 2010, 2011). 
Also the bimodal HB of NGC~1851 is complex, with the RHB stars separated into
two groups. The vast majority are O-rich and Na-poor, while about 10-15\% are
Na-rich  and moderately O-poor. A separate Na-O anti-correlation is seen among
BHB stars. We suggested that most BHB stars descend from the f-SGB stars and
are older and that most RHB stars descend from the b-SGB ones and are younger, 
but the correspondence is  probably not one-to-one. Finally, 47 Tuc and M~5 were 
discussed in Gratton et al. (2013). The cluster 47~Tuc is a
simpler case, with a clear correlation  between the location on the HB and the
Na and O abundances (i.e. like M~4). Instead, while RHB stars in M~5 are
invariably Na-poor and O-rich, the case is more complex for BHB stars, and the
lack of a tight correlation between colours and chemical composition for these
stars requires some additional mechanism to explain observations.

\begin{table*}[htb]\centering
\caption[]{Summary of results on Na-O abundances along the HB in various clusters}
\begin{tabular}{ccc}
\hline
Cluster  & Red HB                             & Blue HB                        \\
\hline
47~Tuc   & [Na/O] correlates with colour      &                                \\
(Gratton et al. 2013) & (increases as B-V decreases)       &                   \\
\hline
NGC~1851 & Mostly O-rich/Na-poor              & Na-rich/O-poor                 \\ 
(Gratton et al. 2012a) & 10-15\% Na-rich, moderately O-poor & (Na-O anticorrelation)         \\
	     & Mostly descendants from b-SGB      & Mostly descendants from f-SGB  \\
\hline
NGC~2808 & O-rich                             & Moderately O-poor              \\
(Gratton et al. 2011) & Spread Na (correlated with colour) & Moderately Na-rich\\
\hline
M~5      & Na-poor, O-rich	                  & Na-O anticorrelation           \\
(Gratton et al. 2013) &                       & Most stars with abundances     \\
         &					                  & similar to RHB stars           \\
\hline
\end{tabular}
\label{t:tabsum}
\end{table*} 

In this paper we focus on M~22 (=NGC~6656), a very intriguing GC. Marino et al.
(2009, 2011) show that there are two populations in this cluster with
different values of [Fe/H], robustly confirming findings based on calcium (Da 
Costa et al. 2009; Lee  et al. 2009). The two populations can be identified
with the two SGBs (Marino et al. 2009, 2012): the metal-poor RGB population is
the descendant of the b-SGB, and the metal-rich one of the f-SGB. Both
populations  display a separate Na-O anti-correlation. The metal-rich population
also appears to be  much richer in $s-$process elements, and Marino et  al.
(2012) suggest that it is  also richer in the sum of the CNO elements, a fact
that could help explain the split SGB since at a given age and metallicity,
stars richer  in CNO elements are fainter on the SGB. These properties of M~22
closely resemble those of NGC~1851, but its predominantly blue HB does not
present the striking bi-modality seen in the latter, maybe because of the
different metal content  ([Fe/H]=-1.70 for M~22 vs [Fe/H]=-1.18 for NGC~1851:
Harris, 1996).  Two recent papers have studied the HB of M~22: a high dispersion
study of seven among the coolest non-variable HB stars of M~22 has been
presented by Marino et al. (2013a), while Salgado et al.  (2013) employed
low-resolution blue spectra to measure masses over a large portion of the HB.
Marino et al. (2013a) found that all these stars are Ba-poor and Na-poor. This
favours the hypothesis that the position of a star along the HB is strictly
related  to the chemical composition and that these stars all belong to the
first (Na-poor and He-normal) portion of the metal-poor population of M~22. This
agrees with an analysis of the whole colour-magnitude diagram by Joo \& Lee
(2013), who identified BHB stars with the metal-poor population and the extreme
BHB stars with the metal-rich one. According to these last authors, a large
difference in helium abundance ($Y=0.23$\ to  0.32) is required to explain the
HB. Our analysis allows extending the study of Marino  et al. (2013a) to a much
larger sample of HB stars.

Section 2 presents the observations and data reduction, explaining how they
differ from what has been done for the other GCs studied in this series. Derivation of
the atmospheric parameters and detail of the abundance analysis are  given in
Section 3. Section 4 presents the results and assignment of the stars to three
different populations based on a statistical  cluster analysis. Section 5
presents a  discussion of the HB of this cluster based on these results and on a
comparison with  evolutionary models. Conclusions are drawn in Section 6.

\begin{center}
\begin{figure}
\includegraphics[width=8.8cm]{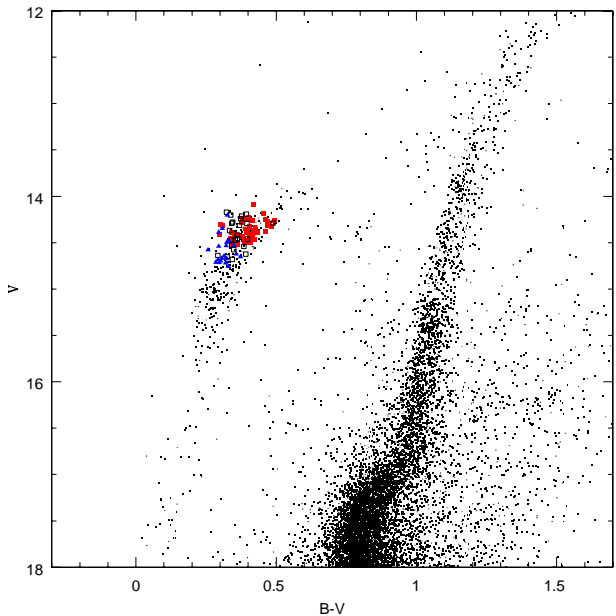}
\caption{($V, B-V$) colour-magnitude diagram of M~22 from Monaco et al. (2004). 
Different symbols are for stars of different groups (see Section 4): Group 1: 
red filled squares; Group 2: black open squares; Group 3: blue filled triangles. 
Dots are the stars not observed in this paper.}
\label{f:fig1}
\end{figure}
\end{center}

\begin{table}[htb]\centering
\caption[]{Observing log}
\setlength{\tabcolsep}{1.5mm}
\begin{tabular}{ccccccc}
\hline
Set up  &    Date     &  UT      &  Exp. time & Airmass & Seeing \\
HR      &             &  Start      & (s) & & (arcsec) \\
\hline             
 3 &  2013-04-25 & 05:27:57 & 2400 & 1.53 & 1.14 \\
 3 &  2013-04-25 & 06:09:42 & 2400 & 1.30 & 1.25 \\
 3 &  2013-07-06 & 07:49:52 & 2400 & 1.47 & 2.19 \\
 3 &  2013-07-10 & 02:56:02 & 2400 & 1.04 & 0.92 \\
 3 &  2013-08-01 & 03:12:40 & 2400 & 1.01 & 0.91 \\
 3 &  2013-08-01 & 04:37:01 & 2400 & 1.12 & 0.77 \\
12 &  2011-09-11 & 03:16:50 & 1500 & 1.41 & 1.29 \\
19A&  2011-06-29 & 07:21:31 & 2500 & 1.21 & 1.84 \\
\hline
\end{tabular}
\label{t:tab0}
\end{table} 

\begin{table*}[htb]
\centering
\caption[]{Photometric data (only available in electronic form)}
\begin{scriptsize}
\begin{tabular}{lccccccccccccc}
\hline
Star& RA (J2000)   & Dec (J2000)  & V      & err   & B      & err   & I      & err   & K      & y      & b      & v      & E(B-V)\\
\hline
1	& 18~36~28.032 & -24~00~26.56 & 14.090 & 0.006 & 14.509 & 0.006 & 13.451 & 0.008 & 12.677 &        &        &        & 0.304 \\
2	& 18~37~03.597 & -23~58~10.33 & 14.182 & 0.006 & 14.638 & 0.008 & 13.483 & 0.008 & 12.884 &        &        &        & 0.359 \\
4	& 18~35~57.058 & -24~04~02.93 & 14.171 & 0.006 & 14.494 & 0.006 & 13.613 & 0.007 & 12.225 &        &        &        & 0.327 \\
5	& 18~36~27.299 & -23~51~44.57 & 14.204 & 0.006 & 14.530 & 0.008 & 13.732 & 0.006 & 13.268 & 14.160 & 14.419 & 14.702 & 0.337 \\
7	& 18~36~20.968 & -24~05~59.01 & 14.193 & 0.006 & 14.587 & 0.006 & 13.558 & 0.006 & 12.943 &        &        &        & 0.327 \\
8	& 18~36~27.716 & -23~52~30.85 & 14.213 & 0.007 & 14.590 & 0.006 & 13.669 & 0.007 &        & 14.184 & 14.453 & 14.765 & 0.342 \\
10	& 18~36~20.551 & -23~58~39.19 & 14.202 & 0.006 & 14.540 & 0.006 & 13.688 & 0.007 & 13.112 & 14.133 & 14.427 & 14.733 & 0.297 \\
11	& 18~36~31.111 & -23~51~45.13 & 14.262 & 0.006 & 14.678 & 0.006 & 13.667 & 0.007 & 13.200 & 14.226 & 14.535 & 14.878 & 0.352 \\
12	& 18~37~08.735 & -23~57~16.79 & 14.327 & 0.006 & 14.810 & 0.006 & 13.627 & 0.008 & 12.921 &        &        &        & 0.355 \\
13	& 18~36~22.368 & -23~59~42.69 & 14.226 & 0.006 & 14.627 & 0.006 & 13.629 & 0.008 & 12.989 &        &        &        & 0.301 \\
14	& 18~36~32.388 & -23~49~35.53 & 14.292 & 0.006 & 14.634 & 0.006 & 13.796 & 0.008 & 13.100 & 14.255 & 14.533 & 14.804 & 0.338 \\
15	& 18~36~16.730 & -23~58~14.50 & 14.231 & 0.007 & 14.634 & 0.007 & 13.683 & 0.007 &        &        &        &        & 0.340 \\
16	& 18~36~29.247 & -23~57~02.07 & 14.240 & 0.007 & 14.614 & 0.007 & 13.742 & 0.007 & 12.914 & 14.152 & 14.452 & 14.779 & 0.341 \\
17	& 18~36~07.706 & -23~55~43.39 & 14.247 & 0.007 & 14.644 & 0.007 & 13.566 & 0.014 & 12.776 & 14.139 & 14.477 & 14.846 & 0.346 \\
18	& 18~36~29.351 & -24~00~28.56 & 14.238 & 0.006 & 14.619 & 0.006 & 13.655 & 0.006 & 13.053 &        &        &        & 0.304 \\
20	& 18~36~18.877 & -23~56~48.84 & 14.250 & 0.006 & 14.711 & 0.006 & 13.593 & 0.007 & 12.804 & 14.167 & 14.535 & 14.909 & 0.339 \\
21	& 18~36~34.560 & -23~54~35.57 & 14.266 & 0.007 & 14.759 & 0.006 & 13.580 & 0.007 & 12.676 & 14.170 & 14.549 & 14.932 & 0.354 \\
23	& 18~36~20.369 & -23~53~05.12 & 14.288 & 0.006 & 14.777 & 0.007 & 13.617 & 0.009 & 12.757 &        &        &        & 0.349 \\
24	& 18~36~21.337 & -23~55~03.24 & 14.277 & 0.008 & 14.746 & 0.009 & 13.647 & 0.018 &        &        &        &        & 0.360 \\
25	& 18~36~32.170 & -23~53~01.18 & 14.292 & 0.007 & 14.688 & 0.007 & 13.744 & 0.010 & 12.981 & 14.210 & 14.532 & 14.861 & 0.346 \\
26	& 18~36~16.749 & -23~53~50.95 & 14.298 & 0.006 & 14.770 & 0.006 & 13.598 & 0.006 &        & 14.256 & 14.621 & 15.012 & 0.354 \\
29	& 18~36~37.359 & -23~59~56.05 & 14.278 & 0.006 & 14.623 & 0.007 & 13.739 & 0.007 & 13.072 &        &        &        & 0.294 \\
31	& 18~36~20.332 & -23~59~37.93 & 14.283 & 0.006 & 14.661 & 0.006 & 13.714 & 0.006 & 13.113 &        &        &        & 0.307 \\
33	& 18~35~55.275 & -24~03~53.28 & 14.302 & 0.006 & 14.604 & 0.006 & 13.741 & 0.007 &        &        &        &        & 0.301 \\
34	& 18~36~17.938 & -23~55~00.89 & 14.309 & 0.006 & 14.777 & 0.007 & 13.630 & 0.009 & 12.884 & 14.226 & 14.595 & 14.991 & 0.349 \\
35	& 18~36~12.781 & -23~55~38.98 & 14.314 & 0.006 & 14.683 & 0.006 & 13.755 & 0.007 &        & 14.258 & 14.543 & 14.884 & 0.324 \\
38	& 18~36~10.313 & -23~59~35.76 & 14.312 & 0.006 & 14.620 & 0.008 & 13.785 & 0.007 & 13.051 &        &        &        & 0.297 \\
39	& 18~36~30.809 & -23~54~25.39 & 14.331 & 0.006 & 14.736 & 0.006 & 13.741 & 0.009 &        & 14.267 & 14.581 & 14.929 & 0.362 \\
40	& 18~36~09.803 & -23~55~13.11 & 14.339 & 0.006 & 14.743 & 0.008 & 13.739 & 0.007 & 13.102 & 14.265 & 14.577 & 14.935 & 0.346 \\
41	& 18~36~23.753 & -23~50~15.98 & 14.384 & 0.006 & 14.781 & 0.006 & 13.802 & 0.006 & 13.100 & 14.351 & 14.650 & 14.970 & 0.336 \\
42	& 18~36~37.681 & -23~57~24.37 & 14.334 & 0.006 & 14.760 & 0.006 & 13.701 & 0.008 & 12.884 & 14.280 & 14.622 & 14.983 & 0.319 \\
43	& 18~36~02.427 & -23~53~38.28 & 14.355 & 0.006 & 14.780 & 0.006 & 13.696 & 0.006 & 12.938 & 14.301 & 14.648 & 15.027 & 0.336 \\
45	& 18~35~50.826 & -23~55~14.43 & 14.350 & 0.006 & 14.739 & 0.008 & 13.740 & 0.007 & 13.032 &        &        &        & 0.336 \\
46	& 18~36~54.722 & -23~56~02.12 & 14.416 & 0.006 & 14.777 & 0.006 & 13.855 & 0.007 & 13.056 &        &        &        & 0.343 \\
47	& 18~36~18.200 & -23~57~44.95 & 14.347 & 0.006 & 14.655 & 0.006 & 13.869 & 0.007 & 13.350 & 14.274 & 14.550 & 14.838 & 0.351 \\
48	& 18~36~36.443 & -23~51~34.51 & 14.387 & 0.006 & 14.822 & 0.008 & 13.784 & 0.007 & 13.117 & 14.348 & 14.671 & 15.007 & 0.335 \\
50	& 18~36~21.568 & -23~55~28.85 & 14.370 & 0.006 & 14.791 & 0.006 & 13.768 & 0.007 & 12.983 & 14.318 & 14.649 & 15.012 & 0.353 \\
51	& 18~36~34.273 & -23~49~31.01 & 14.430 & 0.006 & 14.846 & 0.006 & 13.830 & 0.007 & 13.341 &        &        &        & 0.339 \\
52	& 18~36~23.316 & -23~58~54.13 & 14.369 & 0.006 & 14.703 & 0.008 & 13.856 & 0.007 & 13.308 & 14.320 & 14.591 & 14.900 & 0.307 \\
54	& 18~36~26.414 & -23~55~41.96 & 14.379 & 0.006 & 14.842 & 0.007 & 13.750 & 0.007 & 12.859 & 14.314 & 14.666 & 15.034 & 0.350 \\
55	& 18~36~19.075 & -23~53~16.95 & 14.400 & 0.006 & 14.821 & 0.008 & 13.817 & 0.007 & 12.925 &        &        &        & 0.349 \\
56	& 18~35~55.361 & -23~48~07.38 & 14.459 & 0.006 & 14.795 & 0.006 & 13.889 & 0.006 & 13.311 &        &        &        & 0.347 \\
60	& 18~36~13.546 & -23~53~37.28 & 14.407 & 0.008 & 14.813 & 0.007 & 13.883 & 0.007 & 12.826 &        &        &        & 0.347 \\
61	& 18~36~13.299 & -24~02~24.02 & 14.388 & 0.006 & 14.730 & 0.006 & 13.857 & 0.007 & 13.395 &        &        &        & 0.301 \\
62	& 18~36~15.560 & -23~57~27.25 & 14.389 & 0.006 & 14.684 & 0.006 & 13.913 & 0.006 & 13.336 & 14.329 & 14.584 & 14.867 & 0.341 \\
63	& 18~36~20.844 & -23~56~57.59 & 14.391 & 0.006 & 14.784 & 0.006 & 13.829 & 0.007 & 13.143 & 14.320 & 14.649 & 14.983 & 0.341 \\
67	& 18~36~13.521 & -23~55~08.17 & 14.410 & 0.006 & 14.761 & 0.006 & 13.854 & 0.007 & 13.642 & 14.339 & 14.637 & 14.977 & 0.346 \\
68	& 18~36~38.183 & -23~48~19.48 & 14.478 & 0.006 & 14.869 & 0.008 & 13.893 & 0.007 & 13.404 &        &        &        & 0.339 \\
69	& 18~36~23.533 & -23~56~04.66 & 14.408 & 0.007 & 14.817 & 0.007 & 13.858 & 0.007 & 12.942 & 14.331 & 14.642 & 14.988 & 0.339 \\
70	& 18~36~50.837 & -23~58~04.35 & 14.472 & 0.006 & 14.879 & 0.006 & 13.877 & 0.007 & 13.201 &        &        &        & 0.343 \\
72	& 18~36~25.842 & -23~51~06.89 & 14.454 & 0.006 & 14.817 & 0.008 & 13.952 & 0.007 & 13.018 & 14.419 & 14.694 & 14.996 & 0.346 \\
73	& 18~35~51.258 & -23~51~37.71 & 14.449 & 0.006 & 14.858 & 0.006 & 13.783 & 0.006 & 13.068 &        &        &        & 0.336 \\
76	& 18~35~38.506 & -23~59~41.47 & 14.415 & 0.006 & 14.714 & 0.006 & 13.890 & 0.006 & 13.446 &        &        &        & 0.327 \\
77  & 18~36~40.999 & -23~58~31.27 & 14.484 & 0.006 & 14.890 & 0.006 & 13.896 & 0.007 & 13.241 &        &        &        & 0.343 \\
79	& 18~35~41.235 & -23~55~44.50 & 14.427 & 0.006 & 14.790 & 0.008 & 13.825 & 0.007 & 13.271 &        &        &        & 0.336 \\
80	& 18~36~24.876 & -23~56~42.30 & 14.430 & 0.006 & 14.811 & 0.006 & 13.884 & 0.007 & 13.405 & 14.384 & 14.681 & 15.002 & 0.351 \\
81	& 18~36~01.511 & -23~51~44.26 & 14.470 & 0.006 & 14.893 & 0.006 & 13.828 & 0.007 & 13.266 &        &        &        & 0.352 \\
83	& 18~37~17.808 & -23~52~31.58 & 14.575 & 0.006 & 14.924 & 0.008 & 14.043 & 0.007 & 13.589 &        &        &        & 0.355 \\
84	& 18~36~26.252 & -23~53~17.37 & 14.463 & 0.006 & 14.861 & 0.008 & 13.876 & 0.007 & 10.745 & 14.423 & 14.724 & 15.062 & 0.334 \\
87	& 18~35~41.848 & -23~49~41.49 & 14.515 & 0.006 & 14.856 & 0.006 & 13.922 & 0.006 & 13.400 &        &        &        & 0.347 \\
88	& 18~36~18.315 & -23~54~17.71 & 14.472 & 0.007 & 14.813 & 0.007 & 13.968 & 0.008 & 14.490 &        &        &        & 0.351 \\
89	& 18~35~57.467 & -23~47~59.08 & 14.539 & 0.007 & 14.834 & 0.008 & 14.085 & 0.009 & 13.520 &        &        &        & 0.347 \\
90	& 18~36~31.690 & -23~55~11.47 & 14.471 & 0.006 & 14.891 & 0.006 & 13.871 & 0.007 & 	      & 14.435 & 14.769 & 15.116 & 0.350 \\
91	& 18~36~17.798 & -23~55~37.53 & 14.470 & 0.006 & 14.819 & 0.006 & 13.946 & 0.007 & 13.776 & 14.416 & 14.699 & 15.027 & 0.353 \\
92	& 18~36~07.048 & -23~51~54.35 & 14.500 & 0.006 & 14.909 & 0.006 & 13.835 & 0.006 & 13.361 & 14.435 & 14.763 & 15.113 & 0.341 \\
94	& 18~36~21.227 & -23~49~07.40 & 14.533 & 0.006 & 14.882 & 0.006 & 14.051 & 0.007 & 13.610 & 14.514 & 14.788 & 15.079 & 0.352 \\
95	& 18~36~23.529 & -23~52~33.80 & 14.496 & 0.006 & 14.820 & 0.008 & 14.016 & 0.007 & 14.313 & 14.458 & 14.709 & 14.991 & 0.364 \\
96	& 18~36~26.039 & -23~56~26.15 & 14.473 & 0.006 & 14.827 & 0.006 & 13.962 & 0.007 & 13.505 & 14.429 & 14.721 & 15.019 & 0.341 \\
98	& 18~36~41.823 & -23~56~01.35 & 14.486 & 0.006 & 14.874 & 0.007 & 13.878 & 0.006 & 13.347 & 14.433 & 14.740 & 15.086 & 0.341 \\
100	& 18~36~10.753 & -23~58~14.26 & 14.479 & 0.008 & 14.810 & 0.007 & 14.036 & 0.007 &        &        &        &        & 0.351 \\
102	& 18~36~40.999 & -23~58~31.27 & 14.484 & 0.006 & 14.887 & 0.008 & 13.847 & 0.008 &        &        &        &        & 0.343 \\
104	& 18~36~26.957 & -23~55~51.11 & 14.499 & 0.006 & 14.889 & 0.006 & 13.939 & 0.007 & 13.331 & 14.450 & 14.745 & 15.066 & 0.341 \\
105	& 18~36~29.377 & -23~53~11.04 & 14.532 & 0.007 & 14.854 & 0.007 & 14.075 & 0.009 &        & 14.468 & 14.727 & 15.016 & 0.346 \\
106	& 18~36~21.457 & -23~57~50.18 & 14.519 & 0.006 & 14.882 & 0.006 & 13.983 & 0.007 & 13.406 & 14.461 & 14.770 & 15.081 & 0.323 \\
107	& 18~36~28.185 & -23~56~29.13 & 14.528 & 0.006 & 14.870 & 0.008 & 14.035 & 0.007 &        & 14.481 & 14.744 & 15.032 & 0.341 \\
108	& 18~36~30.649 & -23~54~44.72 & 14.541 & 0.006 & 14.926 & 0.006 & 13.976 & 0.007 & 12.999 & 14.502 & 14.794 & 15.121 & 0.358 \\
110	& 18~36~47.542 & -23~55~49.88 & 14.623 & 0.007 & 14.981 & 0.007 & 14.120 & 0.007 & 13.398 &        &        &        & 0.343 \\
111	& 18~36~28.419 & -23~49~00.28 & 14.624 & 0.006 & 15.016 & 0.006 & 14.091 & 0.007 & 13.345 &        &        &        & 0.338 \\
112	& 18~36~44.423 & -23~52~52.92 & 14.644 & 0.006 & 14.958 & 0.008 & 14.159 & 0.007 & 13.669 &        &        &        & 0.352 \\
113	& 18~36~11.463 & -23~58~26.56 & 14.578 & 0.006 & 14.835 & 0.008 & 14.148 & 0.007 &        &        &        &        & 0.297 \\
115	& 18~36~54.324 & -23~53~15.40 & 14.680 & 0.006 & 14.973 & 0.006 & 14.201 & 0.007 & 13.733 &        &        &        & 0.352 \\
116	& 18~37~14.277 & -23~58~53.31 & 14.712 & 0.006 & 14.997 & 0.006 & 14.271 & 0.007 & 13.827 &        &        &        & 0.359 \\
117	& 18~35~56.442 & -23~54~03.08 & 14.637 & 0.006 & 14.929 & 0.008 & 14.142 & 0.007 & 13.615 &        &        &        & 0.321 \\
118	& 18~36~16.335 & -23~52~25.59 & 14.652 & 0.006 & 14.977 & 0.008 & 14.170 & 0.007 & 13.821 & 14.608 & 14.863 & 15.162 & 0.331 \\
121	& 18~36~25.688 & -23~51~43.41 & 14.673 & 0.006 & 14.996 & 0.007 & 14.226 & 0.007 & 13.705 & 14.627 & 14.882 & 15.141 & 0.337 \\
122	& 18~36~23.919 & -23~55~40.99 & 14.652 & 0.009 & 15.025 & 0.008 & 14.190 & 0.007 & 12.470 &        &        &        & 0.361 \\
123	& 18~36~33.593 & -23~54~59.84 & 14.669 & 0.006 & 14.975 & 0.006 & 14.192 & 0.007 & 13.294 & 14.608 & 14.870 & 15.144 & 0.358 \\
124	& 18~36~28.637 & -23~54~25.19 & 14.681 & 0.007 & 15.025 & 0.007 & 14.199 & 0.008 & 13.008 & 14.598 & 14.877 & 15.180 & 0.353 \\
126	& 18~36~19.377 & -23~48~32.53 & 14.755 & 0.006 & 15.085 & 0.008 & 14.293 & 0.007 & 13.845 &        &        &        & 0.340 \\
127	& 18~36~17.993 & -23~52~33.48 & 14.714 & 0.006 & 15.011 & 0.006 & 14.245 & 0.007 & 13.904 & 14.663 & 14.908 & 15.191 & 0.331 \\
128 & 18~36~33.828 & -23~52~28.28 & 14.716 & 0.006 & 15.036 & 0.006 & 14.270 & 0.007 & 13.400 & 14.674 & 14.928 & 15.196 & 0.337 \\  
129	& 18~36~23.407 & -23~57~17.68 & 14.697 & 0.006 & 14.999 & 0.006 & 14.263 & 0.007 & 12.334 & 14.667 & 14.909 & 15.175 & 0.351 \\
\hline
\end{tabular}
\end{scriptsize}
\label{t:tab1}
\end{table*} 
 
\section{Observations}

The present analysis is based on spectra obtained with the GIRAFFE spectrograph
of the FLAMES multi-object facility at the VLT UT2 Kueyen telescope (Pasquini et al.
2004). FLAMES was used in MEDUSA mode, with individual fibres pointed to stars or 
empty sky positions. Spectra were obtained with three different set ups: HR03 (wavelength
range 4033-4201~\AA, resolution $R\sim 24,800$), HR12 (wavelength range
5821-6146~\AA, resolution $R\sim 18,700$), and HR19A (wavelength range
7745-8335~\AA, resolution $R\sim 13,867$). The HR12 and HR19A set ups were
selected to allow observations of the strongest features of Na~I (the
D resonance doublet) and O~I (the 7771-74~\AA\ high excitation triplet) observable
in metal-poor BHB stars, other lines of the same elements being too weak 
in such stars. High excitation lines of N~I and Mg~II are also present in
these spectral regions. To unambiguously separate the two main populations of M~22 
(the metal-poor one, related to the bright SGB, and the metal-rich one, 
to the faint SGB), we then asked for additional
observing time with the HR03 set up, giving access to several strong Fe~I features and
the resonance line of Sr~II at 4077~\AA, which is the strongest
feature of an element produced by $n-$capture processes. In addition, this spectral
region provides data on Mg~I, Si~II, Ti~II, Fe~II,
and H$\delta$, which could be used to derive a
reddening-free temperature index. This also allowed solving ambiguities related
to differential reddening when analysing the spectra.
The journal of observations is in Table~\ref{t:tab0}.

We focussed our attention on the BHB between the blue edge of the instability strip 
(at an effective temperature of about 7800~K) and the Grundahl jump (at an effective 
temperature of 11,000~K), though our faint limit actually is slightly brighter 
than the Grundahl jump. Cooler stars, most of them RR Lyrae variables, were avoided 
because scheduling their observation at the appropriate phases required for the 
analysis would have been impractical. Hotter stars were also not included because 
their abundances are affected by the impact of microscopic diffusion and radiative 
levitation. When we considered the photometry by Monaco et al. (2004) over 424 bona 
fide HB stars, we counted 28 stars (that is 7\%) with $(B-V)_0>0.15$, hence within 
the instability strip; 221 stars (52\%) with $(B-V)_0<0.15$\ and with $M_V<14.8$, that 
is, within the range of our spectroscopic sample; and 175 stars (41\%) fainter stars. 
Counting stars on the $vby$\ photometry of the cluster from Richter et al. (1999), 
kindly provided by the authors, we found that in their photometry (which covers an area 
of M~22 similar to the one we are considering here) there are 71 stars (that is, 32\% of
total) bluer than the Grundahl jump, 138 stars (61\% of total) in the observed range, 
and 16 redder stars (7\% of total). The small difference between star counts of 
Monaco et al. and Richter et al. most likely depends on different definition of the photometric 
limits. These numbers (and their uncertainties) should be taken into account when 
interpreting our results in terms of the stellar populations of M~22 (see Section 5).

We were able to point fibres on 94 stars, selected from the photometry by Monaco et al.
(2004). The location of these stars on the colour-magnitude diagram is shown in Figure~\ref{f:fig1}. 
The stars were selected to have no neighbour within 2~arcsec that are brighter 
than $M_{\rm target}+2.5$~mag than the target star. One of the stars (\#130 on our notation) was revealed 
to be a field star from its discrepant radial velocity. It also has much stronger lines 
than the stars of M~22. An additional star (\#114 on our notation), though clearly a
cluster member, is a small amplitude (0.04 mag) variable (star KT-51 in the Kaluzny 
\& Thompson 2001 list). When analysed as the other
programme stars are, it yielded odd results, very likely due to large temperature variations during
our observations, which span a few years. All remaining observed stars yielded consistent
radial velocity and are likely members of M~22, whose radial velocity of -146.3~km~s$^{-1}$
(Harris et al. 1996) sets them far from expectations for most field stars.
Fourteen fibres were pointed to empty sky locations.

We collected photometric data for the programme stars from various sources: broad band
$BVI$\ photometry based on data acquired with the WFI camera at the ESO 2.2m
telescope (Monaco et al. 2004); $vby$ Str\"omgren photometry by Richter et al. (1999),
kindly provided by the authors; and 2MASS $JHK$\ photometry (Skrutskie et al.
2006). These photometric data were dereddened using the reddening map by Monaco et al.
(2004). Relevant data are listed in Table~\ref{t:tab1}. While we do not list errors for
individual stars for the $vby$\ photometry, paper, upper limits for the
total photometric errors are 0.015~mag for $V$, 0.019~mag for $(b-y)$, and 0.029~mag 
for $m_1$, according to Richter et al. (1999).

Exposure times were set to provide $S/N\sim 50$ for all the observed spectral
ranges. Observations longer than 45 minutes were then split into several visits that
were performed over several months. Spectra were extracted and calibrated using the 
ESO FLAMES/GIRAFFE pipeline v 2.11.1, running under {\sc GASGANO} environment 
(available at http://www.eso.org/sci/software/pipelines/). Sky subtraction, combination, 
continuum normalization, 
and shifting to rest frame were performed with IRAF\footnote{IRAF is distributed by the 
National Optical Astronomy Observatories, which are operated by the Association of 
Universities for Research in Astronomy, Inc., under cooperative agreement with the 
National Science Foundation.}.
A median of the sky spectra was obtained.
Different spectra for the same star were cross correlated with respect
to the first exposure and brought to common radial velocity before being
combined using the median over different exposures (the spectra have
similar shapes and flux levels); we then applied the
barycentric correction of the first spectrum.
Shift to rest wavelength was done using the radial velocities measured
in the spectra.
A normalisation was done on the spectra using {\sc CONTINUUM} task within IRAF
with a third-order Legendre function.

\begin{table*}[htb]
\centering
\caption[]{S/N of spectra, radial velocities, FWHM, and rotational velocities (only available in electronic form)}
\setlength{\tabcolsep}{1.5mm}
\begin{scriptsize}
\begin{tabular}{lcccccccccccccc}
\hline
Star&S/N(3)&S/N(12)&S/N(19)&RV(3)&RV(12)&RV(19)&$<$RV$>$&r.m.s. &FWHM(3) &FWHM(12)&FWHM(19)&$<$FWHM$>$& rms &$v\sin{i}$\\
    &    &    &    & (km~s$^{-1}$) & (km~s$^{-1}$) & (km~s$^{-1}$) & (km~s$^{-1}$) & (km~s$^{-1}$) & (km~s$^{-1}$) & (km~s$^{-1}$) & (km~s$^{-1}$) & (km~s$^{-1}$) & (km~s$^{-1}$)& (km~s$^{-1}$) \\
\hline
1	& 45 & 75 & 50 & -153.5 & -153.2 & -154.6 &	-153.8 & 0.8 & 21.8 & 18.8 & 22.7 & 21.6 &  2.0 &  7 \\
2	& 67 & 54 & 62 & -149.0 & -142.3 & -144.7 &	-145.3 & 3.4 & 27.7 & 19.5 & 23.6 & 25.4 &  4.1 & 12 \\
4	& 30 & 40 & 43 & -155.5 & -149.8 & -156.2 & -153.8 & 3.5 & 18.7 & 26.3 & 24.9 & 21.5 &  4.0 &  7 \\
5	& 45 & 64 & 49 & -136.1 & -138.7 & -136.7 & -137.2 & 1.4 & 35.1 & 18.9 & 26.5 & 30.3 &  8.1 & 17 \\
7	& 41 & 50 & 50 & -157.9 & -152.9 & -157.5 & -156.1 & 2.8 & 28.2 & 27.8 & 21.8 & 26.3 &  3.6 & 13 \\
8	& 69 & 80 & 65 & -134.4 & -130.4 & -132.5 & -132.4 & 2.0 & 27.6 & 15.2 & 17.0 & 22.8 &  6.7 &  9 \\
10	& 53 & 50 & 49 & -141.8 & -144.4 & -141.0 & -142.4 & 1.8 & 41.0 & 55.1 & 46.6 & 44.6 &  7.1 & 29 \\
11	& 57 & 54 & 55 & -143.1 & -143.4 & -144.9 & -143.8 & 1.0 & 30.5 & 43.0 & 40.8 & 35.2 &  6.7 & 21 \\
12	& 50 & 55 & 44 & -152.6 & -148.4 & -153.6 &	-151.5 & 2.7 & 28.8 & 33.6 & 36.0 & 31.6 &  3.7 & 18 \\
13	& 42 & 52 & 51 & -147.1 & -149.2 & -147.8 & -148.0 & 1.1 & 24.9 & 38.7 & 27.4 & 27.6 &  7.3 & 14 \\
14	& 54 & 50 & 54 & -140.9 & -141.4 & -143.9 & -142.0 & 1.6 & 17.7 & 21.2 & 21.6 & 19.3 &  2.1 &$<$5 \\
15	& 56 & 69 & 51 & -139.8 & -137.2 & -140.1 & -139.1 & 1.6 & 21.9 & 21.2 & 18.2 & 20.8 &  2.0 &  6 \\
16	& 31 & 45 & 61 & -141.9 & -153.4 & -143.0 & -146.1 & 6.4 & 18.9 & 27.8 & 17.5 & 19.8 &  5.6 &$<$5 \\
17	& 	 & 68 & 65 &        & -136.7 & -130.7 & -133.7 & 4.2 &      & 24.6 & 32.3 & 29.8 &  5.5 & 16 \\
18	& 50 & 59 & 50 & -137.1 & -142.2 & -140.8 & -140.0 & 2.6 & 18.4 & 19.2 & 27.6 & 21.1 &  5.1 &  7 \\
20	& 52 & 68 & 63 & -152.4 & -149.9 & -148.6 &	-150.3 & 2.0 & 27.0 & 31.5 & 20.4 & 25.7 &  5.6 & 12 \\
21	& 35 & 56 & 62 & -138.7 & -137.4 & -139.7 &	-138.6 & 1.2 & 21.3 & 25.4 & 19.3 & 21.4 &  3.5 &  7 \\
23	& 52 & 68 & 63 & -147.9 & -142.4 & -145.1 & -145.1 & 2.7 & 33.9 & 28.7 & 32.1 & 32.7 &  2.6 & 19 \\
24	& 56 & 60 & 38 & -143.4 & -141.8 & -147.0 & -144.1 & 2.7 & 33.6 & 23.4 & 30.8 & 31.3 &  5.3 & 18 \\
25	& 36 & 61 & 43 & -132.7 & -135.2 & -133.8 & -133.9 & 1.2 & 28.7 & 17.6 & 18.4 & 24.2 &  6.2 & 11 \\
26	& 42 & 61 & 60 & -157.5 & -151.8 & -155.4 &	-154.9 & 2.9 & 26.2 & 24.9 & 24.8 & 25.6 &  0.8 & 12 \\
29	& 49 & 58 & 56 & -149.3 & -147.7 & -149.2 & -148.7 & 0.9 & 37.3 & 35.9 & 33.4 & 36.0 &  2.0 & 22 \\
31	& 46 & 51 & 51 & -152.0 & -153.3 & -154.1 & -153.1 & 1.0 & 27.2 & 23.0 & 24.8 & 25.9 &  2.1 & 12 \\
33	& 29 & 42 & 37 & -148.9 & -150.8 & -157.2 & -152.3 & 4.4 &      & 62.0 & 57.9 & 25.4 &  2.9 & 12 \\
34	& 58 & 53 & 46 & -148.2 & -147.6 & -146.9 & -147.5 & 0.7 & 31.7 & 28.9 & 34.5 & 32.1 &  2.8 & 18 \\
35	& 59 & 53 & 53 & -151.8 & -149.5 & -150.1 & -150.5 & 1.2 & 27.4 & 28.5 & 24.6 & 26.8 &  2.0 & 13 \\
38	& 57 & 54 & 46 & -157.7 & -154.4 & -154.3 & -155.5 & 1.9 & 31.0 & 28.1 & 27.3 & 29.6 &  2.0 & 16 \\
39	& 34 & 45 & 48 & -126.0 & -138.0 & -125.4 & -129.8 & 7.1 & 49.6 & 31.8 & 47.0 & 46.3 &  9.6 & 30 \\
40	& 43 & 60 & 60 & -154.2	& -145.3 & -150.4 & -150.0 & 4.5 & 18.7 & 21.6 & 24.2 & 20.7 &  2.8 &  6 \\
41	& 49 & 49 & 66 & -151.3 & -146.1 & -150.7 & -149.4 & 2.9 & 19.2 & 17.8 & 23.3 & 20.2 &  2.9 &  5 \\
42	& 36 & 51 & 40 & -138.1 & -137.3 & -135.9 & -137.1 & 1.1 & 33.0 & 23.6 & 23.9 & 29.1 &  5.4 & 16 \\
43	& 60 & 68 & 56 & -150.5 & -148.3 & -150.3 &	-149.7 & 1.2 & 22.3 & 30.2 & 20.5 & 22.9 &  5.1 &  9 \\
45	& 48 & 49 & 54 & -155.5 & -148.3 & -153.4 & -152.4 & 3.7 & 38.5 & 30.3 & 34.0 & 36.0 &  4.1 & 22 \\
46	& 51 & 55 & 51 & -153.2 & -151.1 & -152.8 & -152.4 & 1.1 & 24.2 & 28.9 & 20.3 & 23.8 &  4.3 & 10 \\
47	& 52 & 51 & 48 & -140.4 & -148.5 & -144.5 & -144.5 & 4.0 & 34.2 & 35.1 & 38.0 & 35.4 &  2.0 & 21 \\
48	& 47 & 57 & 67 & -131.4 &        & -139.6 & -135.5 & 5.7 & 34.5 & 26.1 & 37.4 & 34.2 &  5.9 & 20 \\
50	& 50 & 54 & 46 & -140.7 & -140.7 & -137.5 & -139.6 & 1.9 & 29.8 & 34.2 & 22.0 & 28.2 &  6.1 & 15 \\
51	& 37 & 52 & 62 & -134.7 & -145.0 & -133.0 & -137.5 & 6.5 & 45.1 & 33.5 & 28.3 & 38.6 &  8.6 & 24 \\
52	& 44 & 42 & 56 & -150.9 & -150.1 & -146.8 & -149.3 & 2.2 & 75.0 &      & 63.8 & 72.8 &  7.9 & 50 \\
54	& 44 & 50 & 63 & -150.4 & -144.0 & -155.5 & -150.0 & 5.7 & 59.9 & 42.9 & 41.3 & 52.1 & 10.3 & 34 \\
55	& 53 & 55 & 62 & -141.5 & -146.3 & -140.2 & -142.7 & 3.2 & 37.9 & 51.1 & 36.3 & 39.3 &  8.1 & 24 \\
56	& 33 & 	  &    & -146.0 &        & -150.5 & -148.2 & 3.2 & 35.0 & 	   &      & 35.0 &      & 21 \\
60	& 69 & 56 & 51 & -175.0 & -168.7 & -168.5 & -170.7 & 3.7 & 25.9 & 	   & 16.3 & 22.7 &  6.8 &  9 \\
61	& 46 & 51 & 52 & -145.8 & -146.1 & -146.2 & -146.0 & 0.2 & 49.2 & 51.6 & 34.0 & 45.2 &  9.6 & 29 \\
62	& 55 & 63 & 25 & -147.9 & -149.0 & -160.5 & -152.5 & 7.0 & 39.1 & 35.3 & 47.4 & 41.0 &  6.2 & 26 \\
63	& 56 & 63 & 44 & -147.7 & -147.9 & -144.7 & -146.8 & 1.8 & 23.6 & 27.5 & 28.2 & 25.5 &  2.5 & 12 \\
67	& 47 & 47 & 57 & -140.9 & -147.2 & -139.7 & -142.6 & 4.0 & 44.8 & 39.3 & 30.6 & 39.9 &  7.2 & 25 \\
68	& 37 & 46 & 33 & -141.6 & -144.1 & -139.2 & -141.6 & 2.5 & 19.3 & 37.8 & 23.7 & 23.2 &  9.7 &  9 \\
69	& 45 & 56 & 44 & -174.3 & -162.1 & -175.3 & -170.6 & 7.4 & 34.7 & 23.7 & 26.0 & 30.6 &  5.8 & 17 \\
70	& 55 & 56 & 40 & -151.5 & -148.6 & -149.5 & -149.9 & 1.4 & 28.3 & 29.8 & 24.4 & 27.4 &  2.8 & 14 \\
72	& 45 & 52 & 59 & -140.1 & -140.3 & -137.9 & -139.4 & 1.3 & 67.3 & 	   & 45.6 & 60.0 & 15.4 & 40 \\
73	& 44 & 50 & 41 & -148.6 & -142.8 & -142.1 & -144.5 & 3.6 & 35.8 & 23.6 & 41.4 & 35.7 &  9.1 & 21 \\
76	& 38 & 48 & 56 & -133.2 & -152.7 & -137.3 & -141.1 &10.3 & 50.2 & 	   & 50.1 & 50.2 &  0.1 & 33 \\
77	& 52 & 56 & 45 & -154.7 & -148.6 & -154.2 & -152.5 & 3.4 & 29.6 & 33.3 & 24.7 & 28.7 &  4.3 & 15 \\
79	& 55 & 50 & 41 & -155.1 & -152.3 & -164.6 & -157.4 & 6.5 & 50.9 & 32.3 & 52.5 & 48.7 & 11.2 & 32 \\
80	& 43 & 53 & 58 & -141.1 & -146.6 & -134.5 & -140.7 & 6.1 & 62.9 & 	   & 60.1 & 62.0 &  2.0 & 42 \\
81	& 47 & 47 & 52 & -143.3 & -141.9 & -143.1 & -142.7 & 0.8 & 21.9 & 33.6 & 32.1 & 26.5 &  6.4 & 13 \\
83	& 34 & 44 & 39 & -150.8 & -141.7 & -143.0 & -145.2 & 4.9 & 26.0 & 19.9 & 20.9 & 23.7 &  3.3 & 10 \\
84	& 35 & 51 & 60 & -153.7 & -150.2 & -154.6 & -152.8 & 2.3 & 36.1 & 46.0 & 34.5 & 37.0 &  6.2 & 22 \\
87	& 	 & 46 & 55 & 	    &        & -140.0 & -140.0 & 	 &      & 54.5 &      & 54.5 &      & 36 \\
88	& 52 & 56 & 43 & -155.9 & -149.1 & -144.2 & -149.7 & 5.9 & 48.0 & 36.7 & 38.7 & 43.7 &  6.1 & 28 \\
89	& 35 & 41 & 48 & -139.8 & -145.8 & -146.3 & -144.0 & 3.6 & 22.2 & 13.5 & 17.5 & 19.6 &  4.3 &$<$5 \\
90	& 28 & 50 & 46 & -144.0 & -156.3 & -146.2 & -148.8 & 6.6 & 54.8 & 54.8 & 45.0 & 52.0 &  5.6 & 34 \\
91	& 62 & 66 & 46 & -130.7 & -137.0 & -129.8 & -132.5 & 3.9 & 53.1 & 63.0 & 58.1 & 56.0 &  4.9 & 37 \\
92	& 27 & 37 & 36 & -143.7 & -148.5 & -139.1 & -143.8 & 4.7 & 36.1 & 49.8 & 38.5 & 38.7 &  7.3 & 24 \\
94	& 57 & 49 & 53 & -143.4 & -144.5 & -144.6 & -144.2 & 0.6 & 19.0 & 19.0 & 13.4 & 17.4 &  3.3 &$<$5 \\
95	& 37 & 51 & 42 & -135.2 & -144.1 & -137.2 & -138.8 & 4.7 & 21.0 & 24.5 & 18.2 & 20.7 &  3.2 &  6 \\
96	& 54 & 59 & 54 & -147.1 &        & -147.5 & -147.3 & 0.3 & 41.9 & 	   & 29.1 & 37.6 &  9.0 & 23 \\
98	& 32 & 48 & 53 & -148.2 &        & -150.4 & -149.3 & 1.6 & 37.6 & 	   & 27.2 & 34.2 &  7.4 & 20 \\
100	& 49 & 50 & 49 & -132.3 & -133.5 & -137.2 & -134.3 & 2.5 & 21.6 & 23.5 & 20.4 & 21.5 &  1.6 &  7 \\
102	& 40 & 57 & 39 & -164.6 & -161.4 & -161.8 & -162.6 & 1.8 & 29.6 & 23.4 & 15.8 & 24.8 &  7.0 & 11 \\
104	& 50 & 58 & 41 & -160.2 & -160.4 & -158.1 & -159.6 & 1.3 & 33.8 & 24.1 & 29.8 & 31.3 &  4.9 & 18 \\
105	& 36 & 49 & 43 & -150.0 & -140.3 & -144.9 & -145.1 & 4.9 & 38.8 & 37.7 & 39.9 & 38.9 &  1.1 & 24 \\
106	& 52 & 40 & 32 & -146.7 & -148.4 & -154.4 & -149.8 & 4.0 & 66.7 & 50.4 & 72.8 & 66.1 & 11.6 & 45 \\
107	& 40 & 43 & 44 & -147.1 & -143.3 & -142.5 & -144.3 & 2.4 &      & 47.7 & 42.7 & 44.4 &  3.5 & 28 \\
108	& 50 & 53 & 38 & -160.2 & -151.9 & -158.5 & -156.9 & 4.4 & 59.3 & 60.8 & 42.1 & 54.6 & 10.4 & 36 \\
110	& 29 & 47 & 37 & -139.3 & -137.4 & -143.2 & -140.0 & 3.0 & 19.0 & 28.3 & 20.3 & 20.7 &  5.0 &  6 \\
111	& 39 & 36 & 49 & -134.6 & -144.7 & -141.2 & -140.2 & 5.1 &      & 45.9 & 62.4 & 56.9 & 11.6 & 38 \\
112	& 39 & 44 & 38 & -136.1 & -138.5 & -138.9 & -137.8 & 1.5 & 20.3 & 31.8 & 16.4 & 20.8 &  8.0 &  6 \\
113	& 56 & 49 & 47 & -155.3 & -148.9 & -155.7 & -153.3 & 3.8 & 19.6 & 42.9 & 15.3 & 21.7 & 14.9 &  7 \\
115	& 40 & 47 & 39 & -149.4 & -150.6 & -146.0 & -148.7 & 2.4 & 24.9 & 22.0 & 20.2 & 23.1 &  2.4 &  9 \\
116	& 39 & 28 & 24 & -151.3 & -154.7 & -158.1 & -154.7 & 3.4 &      & 13.0 & 18.2 & 16.5 &  3.7 &$<$5 \\
117	& 45 & 44 & 40 & -137.9 & -144.5 & -141.5 & -141.3 & 3.3 & 28.8 & 31.9 & 26.9 & 28.7 &  2.5 & 15 \\
118	& 32 & 42 & 43 & -149.5 & -150.8 & -145.2 & -148.5 & 2.9 & 25.5 & 29.1 & 30.2 & 27.4 &  2.5 & 14 \\
121	& 41 & 49 & 47 & -148.5 & -148.7 & -154.3 & -150.5 & 3.3 & 16.3 & 31.7 & 29.0 & 22.1 &  8.2 &  8 \\
122	& 41 &    &    & -135.4 & -143.6 & 	      & -139.5 & 5.8 & 19.9 & 35.9 & 	  & 23.1 & 11.3 &$<$5 \\
123	& 37 & 45 & 35 & -144.7 &        & -143.0 & -143.8 & 1.2 &      & 	   & 38.4 & 38.4 &      & 24 \\
124	& 44 & 47 & 41 & -139.3 & -134.6 & -139.5 & -137.8 & 2.8 & 17.4 & 21.1 & 19.8 & 18.6 &  1.8 &$<$5 \\
126	& 45 & 47 & 28 & -143.1 & -149.0 & -142.8 & -145.0 & 3.5 & 22.8 & 25.0 & 23.8 & 23.4 &  1.1 & 10 \\
127	& 43 & 50 & 40 & -144.3 & -138.0 & 	      & -141.1 & 4.5 & 23.0 & 	   & 17.4 & 21.2 &  4.0 &  7 \\
128	& 48 & 49 & 45 & -144.3 & -146.7 & -149.4 & -146.8 & 2.6 & 27.6 & 31.2 & 18.5 & 25.5 &  6.5 & 12 \\
129	& 45 & 45 & 45 & -159.4 &        & -161.2 & -160.3 & 1.3 & 17.8 & 	   & 22.7 & 19.4 &  3.5 &$<$5 \\
\hline
\end{tabular}
\end{scriptsize}
\label{t:tab2}
\end{table*} 

\begin{center}
\begin{figure}
\includegraphics[width=8.8cm]{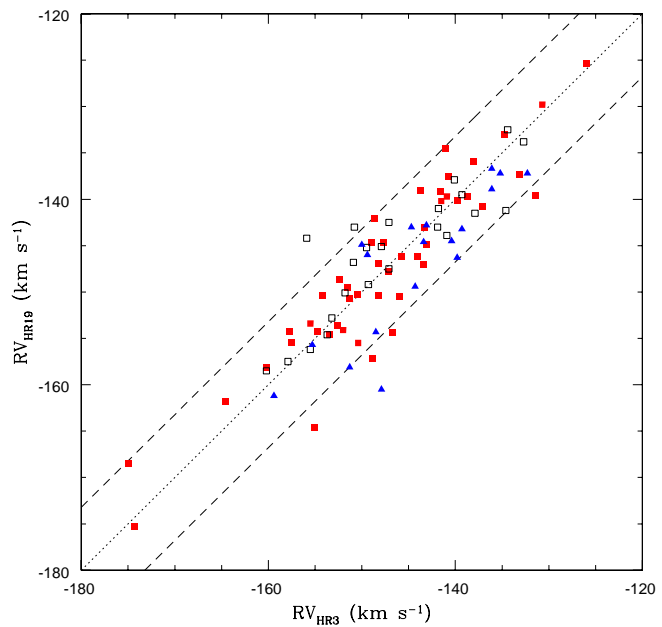}
\caption{Radial velocities from set-up HR03 vs those from set-up HR19A; 
different colours are for stars of different groups (see Section 4): Group 1: 
red filled squares; Group 2: black open squares; Group 3: blue filled triangles.
Dotted line represents equality; dashed lines are $\pm 2$~times the observational
errors.}
\label{f:fig2}
\end{figure}
\end{center}
                  
\subsection{Radial velocities}

We measured radial velocities from the co-added spectra obtained with individual set ups
(see Table~\ref{t:tab2}). They were measured using Gaussian fitting to the position
of three to eight lines in each spectrum and the same lines were used for all stars.
Since very few spectral lines are detectable, radial velocities have quite 
large errors: comparison of different spectra yields typical errors of $\pm 2.4$~km~s$^{-1}$ for 
set-up HR03 and HR19A (see Figure~\ref{f:fig2}), while typical errors are twice as large for set-up HR12. 
Weighted averages have typical errors of $\pm 1.7$~km~s$^{-1}$, fully adequate both as 
a membership criterion and for comparison of the scatter we obtain within different groups of 
stars with the overall dispersion of the radial velocities for the cluster, which we 
measured at $7.7\pm 0.8$~km~s$^{-1}$ for the HB stars we considered. This value (which includes
measurement errors, which are however small) is almost coincident with the value of 
$7.8\pm 0.3$~km~s$^{-1}$ listed by Harris (1996). Since spectra with the HR03 set up 
were obtained two years later than those with the other two set-ups, variations in the 
radial velocity might be used to detect binaries. For two stars (\#88 and \#62), we obtained 
differences in radial velocities of -11.7 and +12.6~km~s$^{-1}$, respectively, when 
comparing data obtained with HR03 and HD19A. These differences are about 3.5 times larger 
(in absolute value) than is typical for other stars: they are then candidate binaries. We 
notice, however, that both of them are quite fast rotators (see next section) and this makes 
their radial velocity measures more uncertain.

We may compare our radial velocities with those measured by Marino et al. (2013a) for 
the four stars in common between the two samples. On average, the difference (ours-Marino et al.)
is $0.3\pm 1.6$~km~s$^{-1}$ (r.m.s.=3.2 km~s$^{-1}$). Most of the scatter is due to star \#17 (for
which we do not have the spectra with the HR03 set-up), which is star \#166 in their 
notation. According to their discussion, this star is suspected of being a blue straggler star 
(BSS). BSS often show radial velocity variations, associated to binarity. If this star is
dropped, on average the difference is $-1.3\pm 0.2$~km~s$^{-1}$, with a very small r.m.s scatter of 
only 0.3~km~s$^{-1}$. We do not attribute much importance to the small zero point difference that
is typical of observations with different set ups of GIRAFFE/UVES. The very small r.m.s.
scatter supports our use of radial velocities for membership, internal dynamics, and
binary detection. In addition, we have 14 stars in common with Salgado et al. (2013). While they
do not claim high accuracy in their velocities, we found excellent agreement with ours, well
within the errors they quoted. On average, our radial velocities are lower by 
$11.1\pm 1.7$~km~s$^{-1}$, with an r.m.s. of 6.3~km~s$^{-1}$\ for individual stars, which is
much less than their quoted error of $\pm 18$~km~s$^{-1}$.

\begin{center}
\begin{figure}
\includegraphics[width=8.8cm]{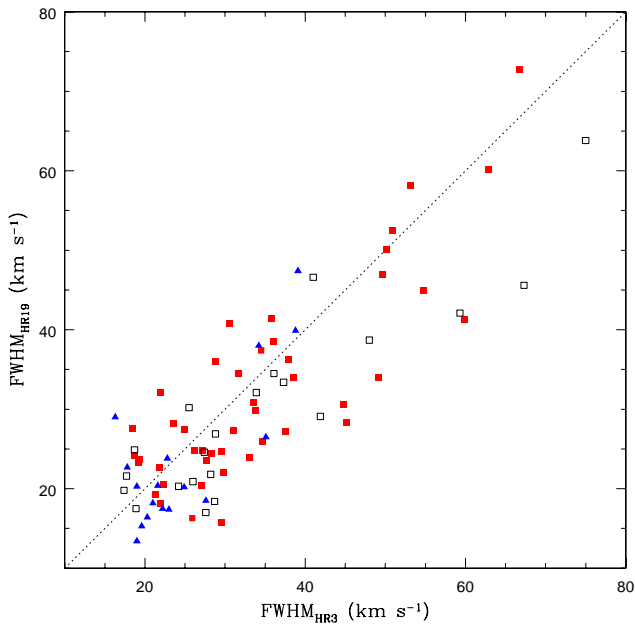}
\caption{FWHM of lines from set up HR03 vs those from set up HR19A; different colours are for stars of
different groups (see Section 4): Group 1: 
red filled squares; Group 2: black open squares; Group 3: blue filled triangles.}
\label{f:fig3}
\end{figure}
\end{center}

\begin{center}
\begin{figure}
\includegraphics[width=8.8cm]{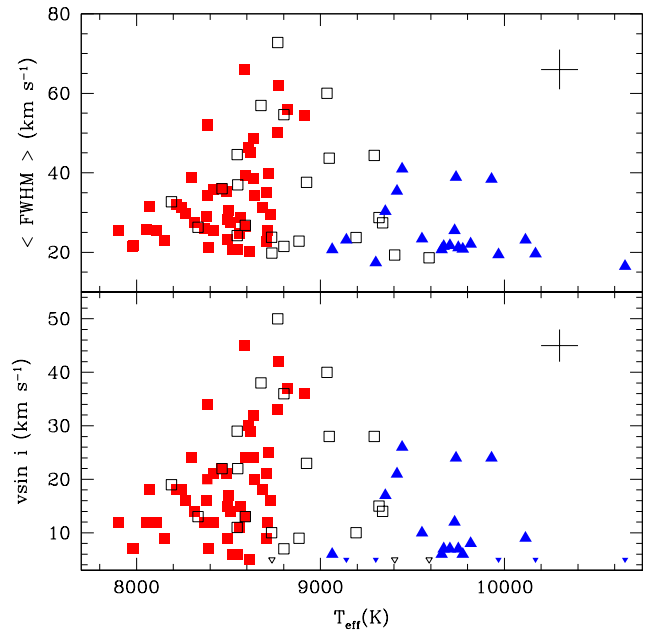}
\caption{\teff vs FWHM of lines (upper panel) and rotational velocity ($v~\sin{i}$: 
lower panel); different colours are for stars of different groups (see Section 4): Group 1: 
red filled squares; Group 2: black open squares; Group 3: blue filled triangles. Small triangles 
in the lower panel represent upper limits in $v~\sin{i}$. Typical error bars are also shown.}
\label{f:fig11}
\end{figure}
\end{center}

\subsection{Rotational velocities}

We found that spectral lines of several stars are 
clearly broader than those of others. This is likely to be due to rotation (see Peterson 
1983; Behr et al. 1999; Recio-Blanco et al. 2004). We then measured the full width 
at half maximum (FWHM) of the spectral lines, which is the convolution of the 
instrumental profile and intrinsic broadening of the spectral lines. The values we 
obtained from individual set ups are listed in Columns 10-12 of Table~\ref{t:tab2}.
The weighted average values (with double weight to results from set ups HR03 and HR19,
where there are stronger lines) are given in Column 13, with their errors in Column 14.
In Figure~\ref{f:fig3} we compare the FWHM of lines from set up HR03 to those from 
set up HR19A. There is quite a good correlation between these measures.
 
An accurate estimate of the errors in line broadening is complex, because it 
depends on line strength (hence temperature), S/N of the spectra, and spectral
resolution, but this is beyond the scope of our work here. An order-of-magnitude estimate can be obtained 
by comparing results obtained with different set ups, after taking 
the different resolving power into account. Typical values are $\pm 5$~km~s$^{-1}$. The lower 
envelope of the distribution (with FWHM$\sim 19$~km~s$^{-1}$) is likely to be populated by 
slowly rotating star, whose profile is dominated by instrumental broadening and turbulent 
motions. FWHM of the rotational broadening may then be obtained by deconvolution of the 
observed FWHM for this value. While our data are not calibrated for this purpose, we 
expect that for profiles dominated by rotational broadening $v~\sin{i}\sim FWHM/\sqrt{2}$. 
We then derived values of $v~\sin{i}$\ using the formula 
$v~\sin{i}=\sqrt{(FWHM^2-FWHM_{\rm ins}^2)/2}$, where the second term takes 
the instrumental profile into account. The values of $FWHM_{\rm ins}$\ we used were those appropriate 
for each set up. Wherever this formula yielded a value of $v~\sin{i}<5$~km~s$^{-1}$, we 
only gave an upper limit of 5~km~s$^{-1}$ to $v~\sin{i}$.
For the four stars in 
common with Marino et al. (2013a), we may compare values of $v~\sin{i}$\ obtained with 
this rough procedure (listed in Column 15 of Table~\ref{t:tab2}) with those they 
obtained from their higher resolution UVES spectra. On average, the difference
(ours-Marino et al. 2013a) is $\Delta v~\sin{i}=-2\pm 3$~km~s$^{-1}$ and the r.m.s. scatter of
6~km~s$^{-1}$\ agrees quite well with our estimate of the errors.

The star with the broadest lines in our sample (\#52) has a FWHM=73~km~s$^{-1}$, 
which corresponds to a rotational velocity of $v~\sin{i}\sim 50$~km~s$^{-1}$. This value is
at the upper limit of the distribution for BHB stars (Peterson et al. 1995; Behr
et al. 2000a, 2000b; Recio-Blanco et al. 2004; Lovisi et al. 2012).

Figure~\ref{f:fig11} shows the run of rotational velocities with effective temperatures. 
At any given temperature, there is some scatter in rotational velocities. The upper envelope 
of the distribution peaks at about 8800 K. The number of stars with evidence of rotation
is a function of temperature: all fast rotators (FWHM$>40$~km~s$^{-1}$) are in the temperature 
range 8400-9400~K.

\begin{table*}[htb]
\centering
\caption[]{Atmospheric parameters (only available in electronic form)}
\begin{scriptsize}
\begin{tabular}{lcccccccccc}
\hline
Star& $T_{\rm eff}(B-V)$&$T_{\rm eff}(V-I)$&$T_{\rm eff}(b-y)$&$T_{\rm eff}(v-y)$&
$T_{\rm eff}(V-K)$&\teff(phot)&err&$T_{\rm eff}(H_\delta)$& $\log{g}$ &	$v_t$ \\
    & (K)& (K)& (K)& (K)& (K)& (K)& (K)& (K)& (dex) & (km~s$^{-1}$)\\
\hline
1	& 8078 & 8102 &      &      & 7489 & 8090 & ~12 & 7987 & 3.10 &	3.61 \\
2	& 8140 & 8132 &      &      & 8372 & 8136 & ~~4 & 7904 & 3.05 &	4.50 \\
4	& 9238 & 8604 &      & 		&      & 8921 & 317 & 8801 & 3.23 &	3.18 \\
5	& 9368 & 9496 & 9471 & 9502 & 9380 & 9468 & ~31 & 9353 & 3.30 &	3.00 \\
7	& 8325 & 8207 &      &      & 8244 & 8266 & ~59 & 8335 & 3.18 &	3.40 \\
8	& 8645 & 8864 & 9314 & 9111 &      & 9009 & 146 & 8883 & 3.24 &	3.22 \\
10	& 8580 & 8642 & 8368 & 8553 & 8502 & 8539 & ~59 & 8547 & 3.25 &	4.74 \\
11	& 8350 & 8583 & 8683 & 8567 &      & 8550 & ~70 & 8489 & 3.20 &	4.86 \\
12	& 8061 & 8117 &      &      & 7977 & 8089 & ~28 & 8072 & 3.15 &	3.40 \\
13	& 8126 & 8220 &      &      & 8049 & 8173 & ~47 & 8313 & 3.23 &	3.46 \\
14	& 9110 & 9252 & 9049 & 9428 & 8534 & 9254 & ~84 & 9404 & 3.36 &	3.12 \\
15	& 8358 & 8809 &      &      &      & 8583 & 225 & 8547 & 3.21 &	3.00 \\
16	& 8677 & 9272 & 8696 & 8676 & 8117 & 8799 & 147 & 8736 & 3.24 &	3.30 \\
17	& 8475 & 8133 & 8300 & 8215 & 7677 & 8267 & ~73 &      & 3.17 &	3.45 \\
18	& 8252 & 8294 &      &      & 8249 & 8273 & ~21 & 8389 & 3.24 &	3.39 \\
20	& 8067 & 8179 & 8122 & 8075 & 7696 & 8104 & ~26 & 8051 & 3.14 &	3.55 \\
21	& 8066 & 8150 & 8123 & 8063 & 7354 & 8093 & ~22 & 7978 & 3.11 &	3.63 \\
23	& 8068 & 8178 &      &      &      & 8123 & ~55 & 8190 & 3.17 &	3.46 \\
24	& 8092 & 8429 &      &      &      & 8261 & 168 & 8244 & 3.16 &	3.45 \\
25	& 8490 & 8867 & 8453 & 8532 & 8210 & 8575 & ~95 & 8547 & 3.23 &	3.31 \\
26	& 8071 & 8112 & 8170 & 8077 &      & 8101 & ~23 & 8106 & 3.15 &	4.00 \\
29	& 8475 & 8457 &      &      & 8089 & 8466 & ~~9 &      & 3.28 &	3.38 \\
31	& 8302 & 8384 &      &      & 8326 & 8343 & ~41 & 8369 & 3.25 &	3.35 \\
33	& 9140 & 8386 &      &      &      & 8763 & 377 & 8712 & 3.31 &	3.20 \\
34	& 8072 & 8153 & 8141 & 8048 &      & 8092 & ~26 & 8218 & 3.18 &	3.46 \\
35	& 8527 & 8580 & 8734 & 8556 &      & 8591 & ~46 &      & 3.27 &	3.24 \\
38	& 8995 & 8556 &      &      & 7932 & 8776 & 219 & 8729 & 3.32 &	3.30 \\
39	& 8561 & 8707 & 8720 & 8565 &      & 8624 & ~43 & 8610 & 3.23 &	3.73 \\
40	& 8409 & 8498 & 8576 & 8407 & 8458 & 8459 & ~41 & 8524 & 3.25 &	4.27 \\
41	& 8378 & 8525 & 8655 & 8701 & 8209 & 8592 & ~72 & 8616 & 3.29 &	3.34 \\
42	& 8102 & 8173 & 8158 & 8130 & 7502 & 8139 & ~16 & 8380 & 3.26 &	3.59 \\
43	& 8183 & 8160 & 8192 & 8110 & 7766 & 8151 & ~18 &      & 3.20 &	3.49 \\
45	& 8447 & 8367 &      &      & 8097 & 8407 & ~40 & 8462 & 3.25 &	3.35 \\
46	& 8885 & 8734 &      &      & 8019 & 8810 & ~76 & 8734 & 3.31 &	3.24 \\
47	&10128 & 9619 & 9306 & 9401 & 9302 & 9571 & 184 & 9416 & 3.34 &	3.50 \\
48	& 8128 & 8396 & 8359 & 8408 & 8246 & 8340 & ~66 & 8385 & 3.26 &	3.38 \\
50	& 8313 & 8547 & 8412 & 8310 & 8023 & 8378 & ~56 & 8494 & 3.25 &	3.40 \\
51	& 8253 & 8444 &      &      & 8886 & 8349 & ~95 & 8637 & 3.32 &	3.19 \\
52	& 8751 & 8751 & 8757 & 8790 & 8691 & 8768 & ~10 &      & 3.34 &	3.15 \\
54	& 8083 & 8360 & 8222 & 8177 & 7549 & 8204 & ~58 & 8386 & 3.24 &	3.58 \\
55	& 8287 & 8634 &      &      & 7691 & 8461 & 173 & 8591 & 3.28 &	3.49 \\
56	& 9364 & 8713 &      &      & 8765 & 9038 & 326 & 8707 & 3.30 &	3.04 \\
60	& 8396 & 9098 &      &      & 7321 & 8747 & 351 & 8708 & 3.29 &	3.55 \\
61	& 8587 & 8565 &      &      & 8862 & 8576 & ~11 & 8620 & 3.34 &	3.19 \\
62	&10194 & 9499 & 9657 & 9602 & 9025 & 9711 & 156 & 9444 & 3.37 &	3.00 \\
63	& 8463 & 8712 & 8345 & 8420 & 8376 & 8472 & ~80 & 8417 & 3.25 &	3.28 \\
67	& 9073 & 8809 & 8787 & 8628 &10023 & 8785 & ~92 & 8719 & 3.30 &	3.50 \\
68	& 8460 & 8534 &      &      & 8938 & 8497 & ~37 &      & 3.30 &	3.14 \\
69	& 8297 & 8783 & 8522 & 8445 & 7632 & 8499 & 102 &      & 3.28 &	3.51 \\
70	& 8348 & 8505 &      &      & 8319 & 8426 & ~79 & 8510 & 3.30 &	3.28 \\
72	& 8893 & 9295 & 9243 & 9196 &      & 9165 & ~90 & 9036 & 3.35 &	4.14 \\
73	& 8283 & 8140 &      &      & 7887 & 8211 & ~72 & 8418 & 3.29 &	3.44 \\
76	& 9738 & 8858 &      &      & 9178 & 9298 & 440 & 8768 & 3.30 &	4.50 \\
77	& 8381 & 8531 &      &      & 8403 & 8456 & ~75 & 8565 & 3.32 &	3.25 \\
79	& 8756 & 8407 &      &      & 8636 & 8581 & 174 & 8634 & 3.31 &	3.64 \\
80	& 8709 & 8947 & 8869 & 8836 &      & 8840 & ~50 & 8770 & 3.31 &	3.15 \\
81	& 8294 & 8314 &      &      & 8623 & 8304 & ~10 & 8595 & 3.31 &	4.25 \\
83	& 9266 & 9120 &      &      & 9377 & 9193 & ~73 &      & 3.42 &	3.00 \\
84	& 8351 & 8481 & 8605 & 8533 &      & 8501 & ~54 & 8551 & 3.32 &	4.34 \\
87	& 9276 & 8548 &      &      & 8873 & 8912 & 364 &      & 3.37 &	4.00 \\
88	& 9344 & 9347 &      & 		&      & 9345 & ~~2 & 9048 & 3.34 &	3.00 \\
89	&10369 & 9858 &      &      & 9195 &10114 & 256 &10169 & 3.52 &	3.00 \\
90	& 8307 & 8531 & 8360 & 8365 &      & 8385 & ~48 &      & 3.27 &	3.35 \\
91	& 9244 & 9175 & 9179 & 8919 &10332 & 9087 & ~72 & 8820 & 3.31 &	3.00 \\
92	& 8319 & 8164 & 8354 & 8332 & 8738 & 8300 & ~43 &      & 3.28 &	4.22 \\
94	& 9225 & 9596 & 9372 & 9401 & 9558 & 9399 & ~76 & 9301 & 3.41 &	3.00 \\
95	&10044 & 9803 &10315 & 9978 &      &10024 & 106 & 9659 & 3.40 &	3.00 \\
96	& 8960 & 9141 & 8827 & 9009 & 9309 & 8989 & ~65 & 8925 & 3.36 &	3.00 \\
98	& 8517 & 8413 & 8594 & 8484 & 9444 & 8498 & ~37 & 8641 & 3.33 &	4.01 \\
100	& 9572 &10052 &      &      & 9444 & 9812 & 240 & 9670 & 3.42 &	3.00 \\
102	& 8365 & 8284 &      &      & 8615 & 8324 & ~41 & 8558 & 3.32 &	3.26 \\
104	& 8485 & 8728 & 8776 & 8767 & 8643 & 8705 & ~69 & 8683 & 3.34 &	3.21 \\
105	& 9650 & 9797 & 9649 & 9537 &      & 9634 & ~53 & 9737 & 3.48 &	3.00 \\
106	& 8599 & 8733 & 8410 & 8593 & 8662 & 8586 & ~66 & 8587 & 3.35 &	3.19 \\
107	& 9141 & 9329 & 9444 & 9440 &      & 9359 & ~71 & 9293 & 3.42 &	4.00 \\
108	& 8751 & 8867 & 9066 & 8890 & 7550 & 8893 & ~65 & 8802 & 3.34 &	3.41 \\
110	& 8924 & 9246 &      &      & 8470 & 9085 & 161 & 9064 & 3.43 &	3.12 \\
111	& 8444 & 8913 &      &      & 8246 & 8679 & 235 & 8676 & 3.39 &	3.32 \\
112	& 9975 & 9572 &      &      & 9386 & 9774 & 202 &      & 3.51 &	3.00 \\
113	&10017 & 9391 &      &      & 9281 & 9704 & 313 &      & 3.55 &	3.00 \\
115	&10593 & 9637 &      &      & 9479 &10115 & 478 &      & 3.56 &	3.00 \\
116	&11097 &10215 &      &      & 9750 &10656 & 441 &      & 3.60 &	3.00 \\
117	& 9760 & 9048 &      &      & 8946 & 9404 & 356 & 9319 & 3.49 &	3.00 \\
118	& 9292 & 9295 & 9459 & 9289 & 9674 & 9325 & ~42 & 9338 & 3.50 &	3.00 \\
121	& 9430 & 9784 & 9576 & 9872 & 9273 & 9707 & 100 & 9817 & 3.55 &	3.00 \\
122	& 8957 & 9982 &      &      &      & 9469 & 513 & 9141 & 3.41 &	3.02 \\
123	&10393 & 9737 & 9818 & 9852 &      & 9930 & 150 &      & 3.53 &	3.00 \\
124	& 9340 & 9613 & 9270 & 9217 &      & 9331 & ~88 & 9591 & 3.53 &	3.00 \\
126	& 9351 & 9651 &      &      & 9494 & 9501 & 150 & 9552 & 3.55 &	3.00 \\
127	& 9890 & 9439 & 9731 & 9605 & 9744 & 9654 & ~95 & 9750 & 3.57 &	3.00 \\
128	& 9487 & 9791 & 9603 & 9762 &      & 9681 & ~71 & 9729 & 3.56 &	2.50 \\
129	&10272 &10180 &10293 &10160 &      &10213 & ~33 & 9968 & 3.53 &	4.50 \\
\hline
\end{tabular}
\end{scriptsize}
\label{t:tab3}
\end{table*} 

\section{Analysis}

\begin{center}
\begin{figure}
\includegraphics[width=8.8cm]{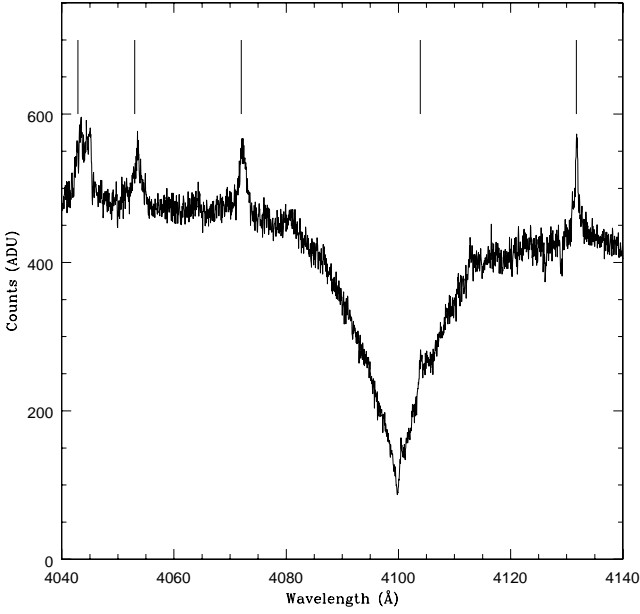}
\caption{Portion of the spectrum of star 115 including $H_\delta$. This is the most extreme case of 
stronger contamination by Th-Ar lines. Identification of several contaminating lines is from the Kitt Peak 
Th-Ar spectrum (http://old-www.noao.edu/kpno/specatlas/thar/ )}
\label{f:contamination}
\end{figure}
\end{center}

\begin{center}
\begin{figure}
\includegraphics[width=8.8cm]{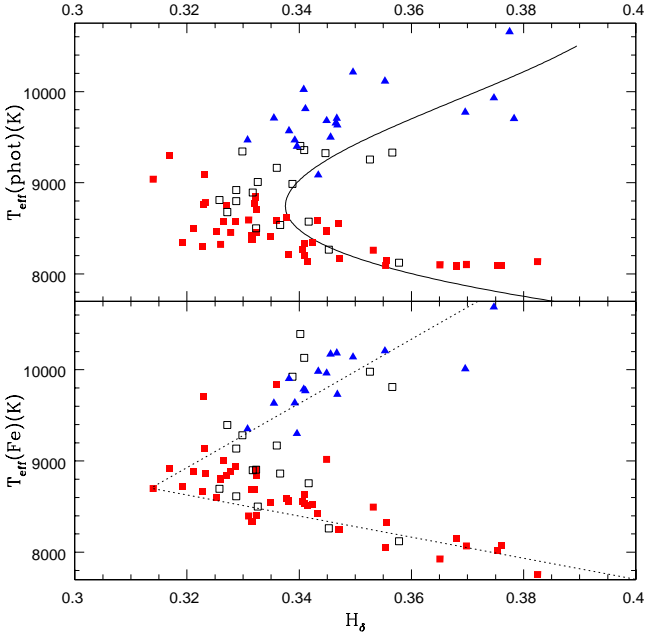}
\caption{Upper panel: Comparison between temperatures from photometry (\teff(Phot)) 
and the index of the strength of the H$\delta$ line ($H\delta$); overimposed
is the relation expected from theoretical models (see Munari et al. 2005). Lower panel: the
same, but for temperatures from Fe lines (\teff(Fe)). Different symbols are for stars of
different groups (see Section 4): Group 1: red filled squares; Group 2: black open 
squares; Group 3: blue filled triangles. The fit lines used to define $T_{\rm Cool}$\ and 
$T_{\rm Hot}$\ are also plotted in the bottom panel.}
\label{f:tfehd}
\end{figure}
\end{center}

\begin{center}
\begin{figure}
\includegraphics[width=8.8cm]{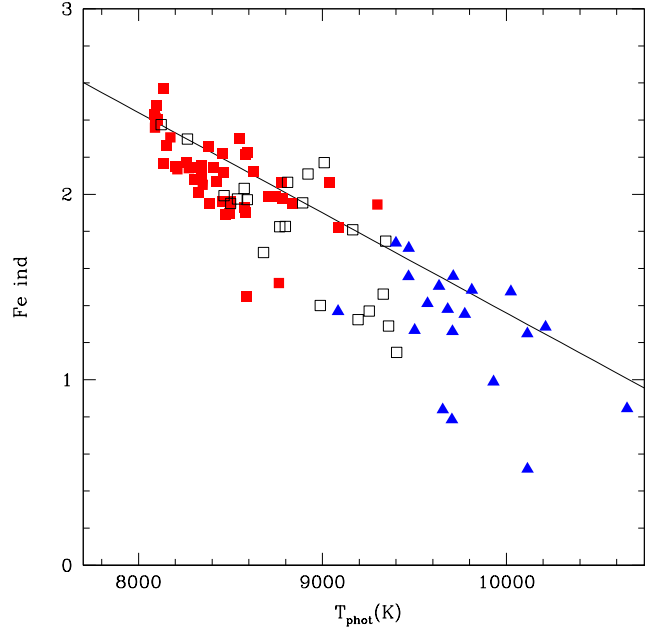}
\caption{Comparison between the Fe I line strength index Fe-Ind and temperatures from 
photometry (\teff(Phot)). Different symbols are for stars of
different groups (see Section 4): Group 1: red filled squares; Group 2: black open 
squares; Group 3: blue filled triangles. Superimposed is the calibration line we used.}
\label{f:tcolfeind}
\end{figure}
\end{center}

\begin{center}
\begin{figure*}
\includegraphics[width=18cm]{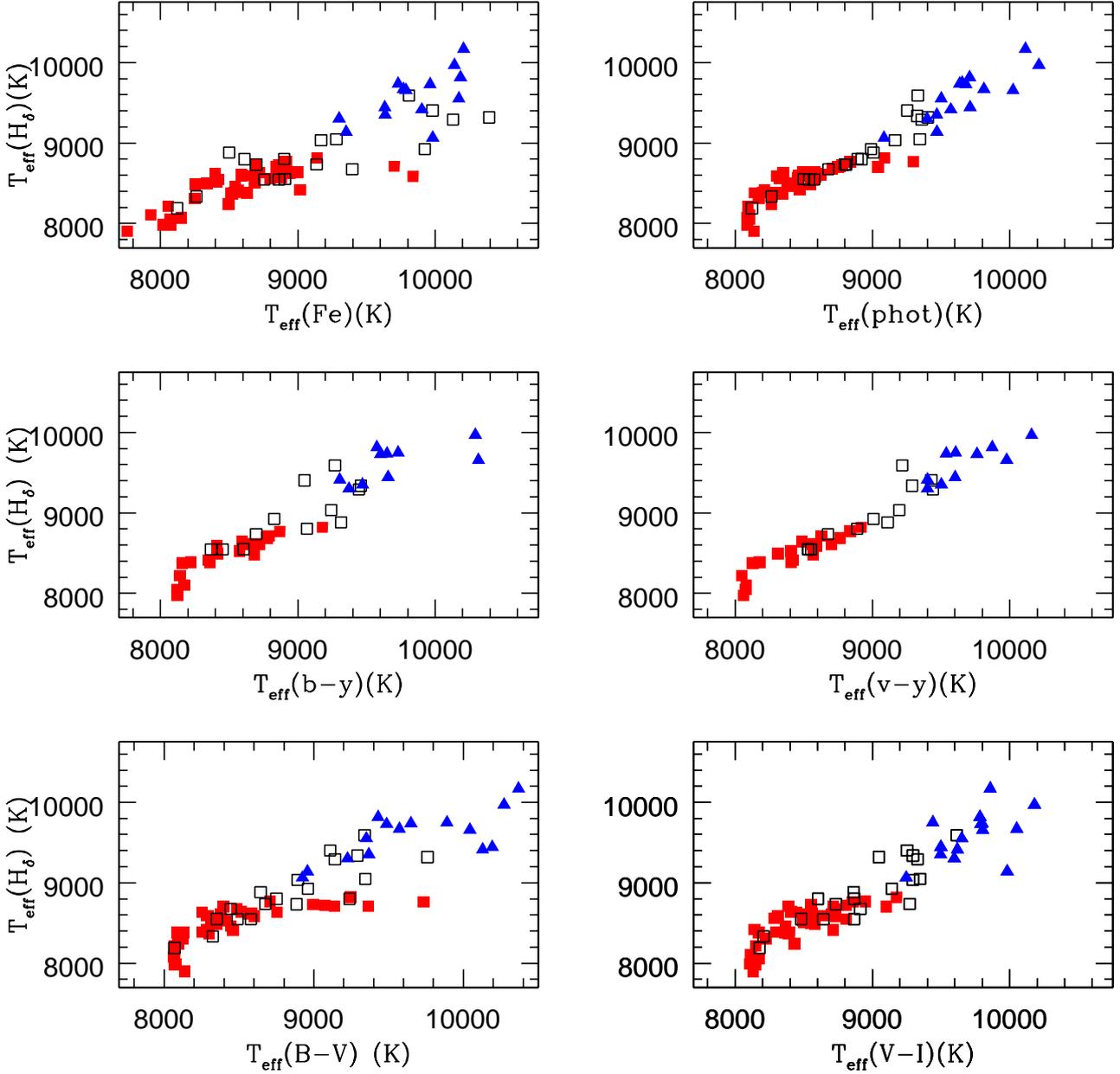}
\caption{Comparison between temperatures from colours and their average (\teff(phot)) and from
our calibration of the H$\delta$ line ($T_{\rm eff}(H\delta)$). Different colours are for stars of
different groups (see Section 4). Group 1: 
red filled squares; Group 2: black open squares; Group 3: blue filled triangles.}
\label{f:fig4}
\end{figure*}
\end{center}

\subsection{Atmospheric parameters}

Our analysis is based on model atmospheres extracted by interpolation within the Kurucz 
(1993) grid. Interpolations were done as described in Gratton \& Sneden (1987) and 
used in many other papers. The grid of models used for this interpolation does not include 
any alpha enhancement. However, at this high temperature, most of the electrons are provided by
hydrogen, not metals. The impact of modifying model metal abundances is then very
small, as confirmed by detailed calculations (see Section 3.5).

The most critical parameter in our abundance analysis is the effective temperature 
\teff. In previous papers of this series (Gratton et al. 2011, 2012a, 2013),
\teff's were derived from colours using calibrations that, for the BHB stars,
were based on Kurucz (1993) model atmospheres. Since red-infrared colours saturate 
for such warm stars, most useful information is provided by visual and near-ultraviolet
colours. Unfortunately, while BVI data were available for all stars, violet
colours were only available for about half of them, and we have no reliable UV photometry
for a significant number of stars.

A basic problem in deriving colours for the programme stars is the variation in
interstellar reddening over the field of M~22. M~22 is seen in projection against the
Galactic bulge and has a high (E(B-V)=0.34: Harris 1996) and differential
reddening (e.g. Richter et al. 1999). A map of such variation has been
prepared by Monaco et al. (2004) covering the whole field of interest. We then corrected
our colours first for the average reddening of M~22 ($E(B-V=0.34$: Harris, 1996) and
then for the differential reddening provided by Monaco et al. (2004). We then derived
temperatures from $B-V$, $V-I$, $b-y$, and $v-y$ colours, reducing them to
a consistent scale, which is the one defined by $B-V$\ colours. We also considered
$V-K$\ colours, but later discarded them because errors were too large
to be useful. We then made a weighted average of these temperatures, assigning
double weight to $v-y$ colours. We called 
\teff(phot) these estimates of the effective temperatures.

These photometric temperatures still contain non-negligible errors, owing not only to
errors in the calibration and photometry of individual stars, but also to uncertainties
in the differential reddening map, which are of the order of 0.01~mag in $E(B-V$), which
corresponds to several hundred K for BHB stars.
Luckily, our spectra offer the opportunity to derive effective temperatures from the
strength of $H_\delta$. Hydrogen lines in BHB stars are also sensitive to gravity (see e.g.
discussion in Marino et al. 2013a); however, this effect does not cause a large
scatter in the relation between effective temperature and strength of the line
because the spread in mass and radius (and then surface gravity) at a given 
\teff\ is actually very small. We then computed an index of the strength of
$H_\delta$ (which we called by this same name) that is the ratio of the flux within 
a region 8~\AA\ wide centred on the line and of the average in two similar reference 
regions located symmetrically with respect to the line at 40~\AA\ separation. We then 
plotted this $H_\delta$ index against \teff(phot) (see Figure~\ref{f:tfehd}).
For a few stars, contamination of the spectra by the wavelength calibration
lamp falsifies this $H_\delta$ index, while the spectra can still be used, with some
care, for other purposes.  
This problem was found for nine stars. Figure~\ref{f:contamination} shows a portion of the
spectrum of star 115 including $H_\delta$, which is the case of strongest contamination.

The parameter $H_\delta$ shows a smooth run with \teff(phot), with a minimum (i.e. strongest
line) at about 8800~K. For comparison, we also plotted the observed relation with a calibration 
based on theoretical models in the upper panel of 
Figure~\ref{f:tfehd}; for this purpose, we used the same definition of the indices to measure
$H_\delta$ indices on the theoretical spectral library by Munari et al. (2005). 
We interpolated the values of the indices we obtained for the observed run of gravity with
temperature. On the whole, there is quite good agreement, though measured $H_\delta$ index
are slightly lower (that is, the line appears stronger) than given by the models. The difference
is small and can be attributed to problems in how the continuum normalization was done on our 
spectra. We attribute the scatter around a mean relation to the effects
of residual differential reddening. To support this claim, we constructed another spectroscopic
index Fe-Ind that we designated as the logarithm of the sum of the equivalent widths of the three strongest Fe~I lines
observable in our spectra (at 4045.82, 4063.60, and 4071.75~\AA), and also plotted this
quantity against \teff(phot) (see Figure~\ref{f:tcolfeind}). We could then fit a
straight line through the observed points and construct a temperature index that we
called \teff(Fe). Plotting $H_\delta$ against \teff(Fe) (see Figure~\ref{f:tfehd}), 
we find that the scatter in the plot is considerably reduced. This is
precisely what we expect if a significant source of scatter in the $H_\delta - T_{\rm eff}$(phot)
plot is due to differential reddening that was not properly taken into account.

We then decided to derive temperatures from the $H_\delta$ index. We prefer to use $H_\delta$
because we might expect some star-to-star variation in Fe abundances. These were obtained
by fitting two straight lines on the $H_\delta - T_{\rm eff}$(Fe) plot: one for stars with
\teff(Fe)$<8500$~K, which we called $T_{\rm Cool}$, and one for stars with \teff(Fe)$>9000$~K,
which we called $T_{\rm Hot}$\ (see Figure~\ref{f:tfehd}). The finally adopted
temperature from H$_\delta$\ ($T_{\rm eff}(H\delta)$) was $T_{\rm Cool}$ for
\teff(phot)$<8400$~K; $T_{\rm Hot}$ if \teff(phot)$>9750$~K; and
$T_{\rm eff}(H\delta)=w*T_{\rm Cool}+(1-w)*T_{\rm Hot}$ if $8400<T_{\rm eff}$(phot)$<9750$~K. In this
last formula, $w=(T_{\rm eff}$(phot)$-8400)/(9750-8400)$. For those stars for
which no $H_\delta$ index could be derived owing to the contamination by the Th-lamp
lines, we adopted \teff(phot) as best estimates of the effective temperatures. 

Figure~\ref{f:fig4} compares $T_{\rm eff}(H\delta)$ with \teff(phot) (as well as with temperatures
from individual colours). The correlation is quite tight, with
an r.m.s. of 144 K. The small residual scatter may be explained by errors in the differential reddening
estimates of $\sim 0.01$~mag, which is well within the accuracy of the method devised by
Monaco et al. (2004). While such corrections appear small, there was clear improvement
on the results from use of $T_{\rm eff}(H\delta)$ rather than \teff(phot). We hence assign
an error of $\pm 100$~K to $T_{\rm eff}(H\delta)$.

Once effective temperatures and differential reddening values were accurately determined, 
surface gravities $\log{g}$\ can be determined from the location of the stars in the 
colour-magnitude diagram (after correction for differential reddening) with very 
small errors, because masses of the stars cannot be 
very different from an average value of 0.63~M$_\odot$ (see Gratton et al. 2010). To this 
purpose, we adopted bolometric corrections from Kurucz (1992, for the metallicity 
of [Fe.H]=-1.70 given by Harris 1996) and a distance modulus of 
$(m-M)_V=13.6$ (Harris, 1996). Surface gravities have errors not larger than $\pm 0.05$~dex.

More critical is the derivation of microturbulent velocities $v_t$. For \& Sneden (2010)
have shown that the value of $v_t$ changes systematically with 
temperature along the HB, reaching a maximum near the RR Lyrae instability strip.
Quite high values of $v_t$ are then appropriate for the cooler stars in our sample, while
lower ones are more appropriate for hotter stars. Our limited spectral range - hence
line list - does not in general allow derivation of reliable $v_t$\ values, so we 
adopted $v_t=3.0$~km~s$^{-1}$ for stars with $T_{\rm eff}(H\delta)>9000$~K, and
$v_t=3.0 - 0.6 (T_{\rm eff}(H\delta)-9000)$\ for cooler stars. However, in about 20\% of the 
cases we had to modify this value, by as much as 1.5~km~s$^{-1}$\ in the most extreme cases, 
to reduce the scatter in abundances from individual lines of O~I and
Fe~I. Errors in these estimates of $v_t$ are quite large, so we think a value of $\pm 1$~km~s$^{-1}$
is appropriate.

Finally, we adopted the same model metal abundance of [A/H]=-1.70 (Harris 1996) for all stars.
While M~22 is known to have a spread in [Fe/H], this is not greater than 0.25 dex peak to
valley (Marino et al. 2011b). An error bar of $\pm 0.2$~dex should then be appropriate for
[A/H].

We may compare our estimate for temperature and gravity with Salgado et al. (2013) ones for the
14 stars in common between the two samples. On average, there are no systematic differences:
the offset (in the sense ours-Salgado et al.) is $93\pm 117$~K (r.m.s.=436~K) for temperatures and 
$0.05\pm 0.07$~dex (r.m.s.=0.27~dex) for gravities.

\begin{center}
\begin{figure}
\includegraphics[width=8.8cm]{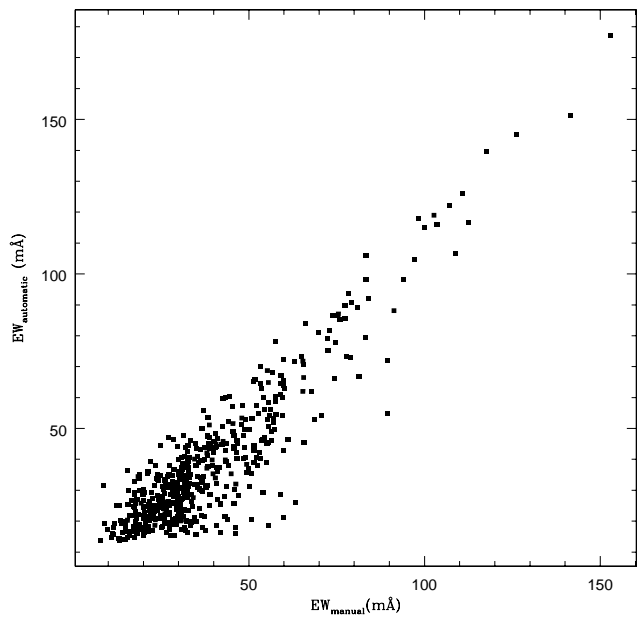}
\caption{Comparison between EWs measured manually and by the automatic procedure described in the
text. The automatic EWs were corrected for the relation given in the text before being plotted.}
\label{f:compew}
\end{figure}
\end{center}

\begin{center}
\begin{figure}
\includegraphics[width=8.8cm]{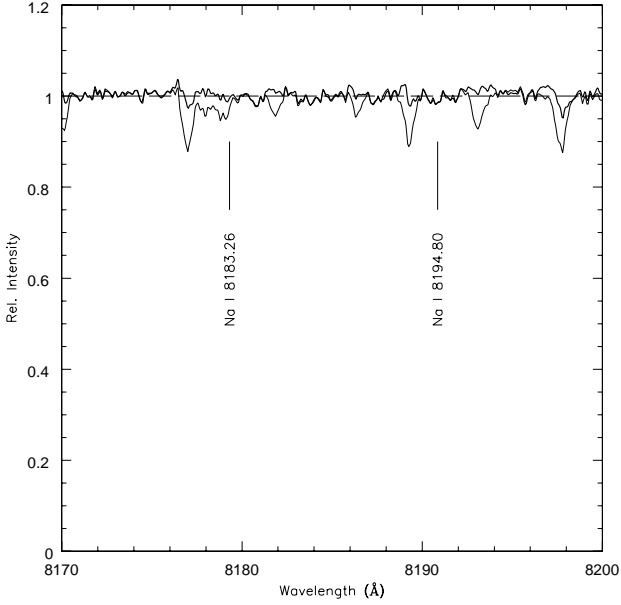}
\caption{A portion of the spectrum of star \#2 including the Na I doublet at 8183-94~\AA before 
(thin line) and after (thick line) the division for telluric lines. Dashed line is an
approximate reference continuum. The Na I lines are very faint, The weakest blue line is not detected;
the strongest red one is at the limit of detection.}
\label{f:telluric}
\end{figure}
\end{center}

\subsection{Equivalent widths}

Our abundances rest on analysis of equivalent widths ($EW$). They were obtained by line 
integration, and are the average of a manual measure, where line edges and local continuum 
level were set by eye inspection of the spectra, and of an automatic measure that is
a measure of the average of the flux within a band four times the FWHM wide centred on the line
(taking the radial velocity of the star into consideration), divided for the average 
fluxes in two comparison ``continuum" regions (each $\sim 2$~\AA\ wide) on both sides of each 
line. The EWs measured with the automatic
procedure were typically slightly lower than those measured manually: 
$EW_{\rm Auto}=0.872~EW_{\rm Manual} - 6.7$~m\AA, with an r.m.s. of the differences of 9.7~m\AA.
The two sets of EWs were put on a uniform scale by correcting the automatic measures to the 
manual ones. Figure~\ref{f:compew} compares the two sets of EWs.

Care was taken to
consider the star-to-star variations in the widths of the lines due to stellar rotation
(see previous paragraph). On the other hand, identification of local continuum and line
blending are generally not a problem, since very few detectable lines are typically present.
In the near infrared, subtraction telluric line subtraction is an issue for N and Na lines. It was 
obtained by dividing the spectra for the average of early type stars with very different radial
velocities obtained throughout our programme (see Figure~\ref{f:telluric}).

Using the Cayrel (1988) formula, we find that the equivalent widths have errors of
$\pm 2$, $\pm 3$, and $\pm 5$~m\AA\ for HR03, HR12, and HR19A spectra, respectively. Errors
are up to twice as large for rapidly rotating stars.

\begin{table*}[htb]
\centering
\caption[]{Abundances for individual stars (only available in electronic form)}
\begin{scriptsize}
\begin{tabular}{lccccccccccccc}
\hline
Star &Y$_{NLTE}$&err&
$[$Fe/H$]_I$ &$[$Fe/H$]_{II}$&$<[$Fe/H$]>$&$[$N/Fe$]$&$[$O/Fe$]$&$[$Na/Fe$]$&$[$Mg/Fe$]_I$&
$[$Mg/Fe$]_{II}$&$[$Si/Fe$]_{II}$&$[$Ti/Fe$]_{II}$&$[$Sr/Fe$]_{II}$\\
\hline
\multicolumn{14}{c}{Group A}\\
1	&      &      & -1.73 & -1.91 & -1.82 & 0.74 & 0.71 & ~0.13 & 0.73 & 0.36 & 0.58 & 0.44 & -0.37 \\
2	&      &      & -1.66 & -1.72 & -1.69 & 0.53 & 0.42 & -0.16 & 0.72 & 0.28 & 0.48 & 0.40 & -0.20 \\
11	&      &      & -1.66 & -1.77 & -1.72 & 0.46 & 0.60 & ~0.12 & 	   & 0.26 & 0.38 & 0.51 & -0.23 \\
12	&      &      & -1.79 & -1.83 & -1.81 &      & 0.83 & -0.01 & 0.74 & 0.52 & 0.80 & 0.43 & -0.47 \\
13	&      &      & -1.67 & -1.75 & -1.71 & 0.74 & 0.71 & -0.01 & 0.88 & 0.49 & 0.63 & 0.42 & -0.55 \\
15	&      &      & -1.63 & -1.91 & -1.77 &      & 0.64 & ~0.02 & 1.10 & 0.81 & 0.48 & 0.36 & -0.44 \\
17	&      &      &       &       &       & 0.32 & 0.49 & -0.01 & 	   &-0.06 &      &      &       \\
18	&      &      & -1.95 & -2.01 & -1.98 & 0.74 & 0.61 & ~0.01 & 0.51 & 0.28 & 0.53 & 0.41 & -0.63 \\
20	&      &      & -1.74 & -1.88 & -1.81 & 0.53 & 0.53 & ~0.01 & 0.29 & 0.76 & 0.50 & 0.34 & -0.28 \\
21	&      &      & -1.78 & -1.89 & -1.84 & 0.69 & 0.60 & -0.19 & 0.82 & 0.39 & 0.43 & 0.46 & -0.49 \\
24	&      &      & -2.01 & -2.07 & -2.04 &      & 0.40 & -0.07 & 	   & 0.77 &      & 0.16 & -0.84 \\
26	&      &      & -1.61 & -1.69 & -1.65 &      & 0.36 & -0.07 & 0.66 & 0.25 & 0.32 & 0.49 & -0.11 \\
31	&      &      & -1.95 & -2.11 & -2.03 &      & 0.86 & -0.05 & 	   & 0.14 & 0.22 & 0.20 & -0.63 \\
33	&      &      & -1.95 & -1.52 & -1.73 &      & 0.71 & -0.07 & 	   & 0.44 & 0.92 &      &       \\	
34	&      &      & -1.60 & -1.82 & -1.71 & 0.44 & 0.49 & -0.09 & 0.83 & 0.35 & 0.55 & 0.33 & -0.26 \\
38	&      &      & -1.75 & -1.94 & -1.85 & 0.64 & 0.72 & -0.13 &      & 0.24 & 0.57 & 0.36 & -0.45 \\
39	&      &      & -1.73 & -1.96 & -1.84 &      & 0.86 & -0.05 & 	   & 0.08 & 0.61 & 0.49 & -0.02 \\
40	&      &      & -1.70 & -1.91 & -1.80 & 0.65 & 0.71 & -0.04 & 0.81 & 0.71 & 0.15 & 0.56 & -0.46 \\
41	&      &      & -1.54 & -1.60 & -1.57 & 0.69 & 0.83 & -0.06 & 0.86 & 0.93 & 0.58 & 0.49 & -0.47 \\
42	&      &      & -1.90 & -2.06 & -1.98 &      & 0.58 & -0.29 & 	   & 0.88 & 0.34 & 0.21 & -0.63 \\
43	&      &      & -1.86 & -1.81 & -1.83 &      & 0.56 & -0.12 & 	   & 0.27 & 0.41 & 0.48 & -0.52 \\
45	&      &      & -1.83 & -1.86 & -1.85 & 0.52 & 0.69 & -0.16 & 	   &      & 0.29 & 0.20 & -0.43 \\
48	&      &      & -2.01 & -2.02 & -2.01 & 0.72 & 0.63 & -0.23 & 	   & 0.74 & 0.65 & 0.23 & -0.68 \\
50	&      &      & -1.68 & -1.78 & -1.73 & 0.71 & 0.61 & ~0.02 & 	   & 0.39 & 0.54 & 0.30 & -0.38 \\
51	&      &      & -1.82 & -1.99 & -1.91 & 0.64 & 0.75 & ~0.03 & 	   & 0.03 &      & 0.35 & -0.31 \\
54	&      &      & -1.94 & -2.03 & -1.98 & 0.52 & 0.67 & ~0.04 & 0.59 & 0.14 &      &-0.06 & -0.32 \\
55	&      &      & -1.79 & -1.99 & -1.89 & 0.59 & 0.61 & ~0.08 & 	   & 0.36 & 0.27 & 0.33 & -0.59 \\
56	&      &      & -1.70 & -1.47 & -1.58 &      &      & -0.06 &      &      &      & 0.53 & -0.23 \\
60	&      &      & -1.87 & -1.83 & -1.85 &      & 0.67 & ~0.02 &      & 0.52 & 0.28 & 0.32 & -1.01 \\
61	&      &      & -2.06 & -1.81 & -1.94 & 0.54 & 0.72 & -0.21 & 	   & 0.88 & 0.15 & 0.00 & -0.73 \\
63	&      &      & -2.16 & -1.85 & -2.00 & 0.82 & 0.74 & -0.09 & 0.55 & 0.52 & 0.20 & 0.14 & -1.02 \\
67	&      &      & -1.85 &       & -1.85 &      & 0.67 & -0.02 &      &      & 	 & 0.19 & -0.54 \\
68	&      &      & -2.08 & -1.70 & -1.89 & 0.89 & 0.65 & -0.26 & 	   & 0.29 & 0.32 & 0.44 & -0.90 \\
69	&      &      & -1.99 & -2.14 & -2.07 & 0.86 & 0.53 & -0.16 & 0.87 &      & 0.52 & 0.35 & -1.30 \\
70	&      &      & -1.96 & -1.91 & -1.94 & 0.76 & 0.75 & -0.20 & 0.48 & 0.89 & 0.50 & 0.33 & -0.63 \\
73	&      &      & -1.91 & -1.75 & -1.83 &      & 0.69 & -0.22 & 	   & 0.09 & 0.51 & 0.34 & -0.68 \\
76	&      &      & -1.91 & -1.69 & -1.80 & 0.66 & 0.68 & -0.09 & 	   & 0.83 & 0.52 &      & -0.43 \\
77	&      &      & -1.99 & -1.89 & -1.94 & 0.79 & 0.74 & ~0.04 & 0.96 &      & 0.48 & 0.34 & -0.75 \\
79	&      &      & -2.04 &       & -2.04 & 0.63 & 0.81 & -0.19 & 	   & 0.84 & 0.75 & 0.11 & -0.66 \\
80	&      &      & -1.85 & -1.93 & -1.89 &      & 0.75 & ~0.02 & 	   & 1.00 &      & 0.33 & -0.28 \\
81	&      &      & -1.88 & -1.69 & -1.78 &      & 0.50 & -0.05 & 	   & 0.78 & 0.40 & 0.29 & -0.44 \\
87	&      &      &       &       &       &      & 0.65 & ~0.02 & 	   & 0.55 & 	 &      &       \\
90	&      &      & -2.09 & -2.18 & -2.14 & 0.63 & 0.65 & -0.15 & 	   & 0.65 & 0.99 & 0.29 & -0.67 \\
91	&      &      & -1.91 & -1.83 & -1.87 & 0.93 & 0.53 & -0.02 & 	   & 0.74 & 0.58 & 0.43 & -0.56 \\
92	&      &      & -2.09 &       & -2.09 & 0.67 & 0.69 & -0.32 & 	   & 0.68 & 0.76 & 0.55 & -1.08 \\
98	&      &      & -1.97 & -1.76 & -1.86 & 0.85 & 0.59 & ~0.07 & 	   & 0.08 & 0.35 & 0.59 & -0.50 \\
102	&      &      & -1.97 & -2.09 & -2.03 & 0.92 & 0.41 & -0.29 & 0.72 & 0.29 & 0.35 & 0.34 & -0.72 \\
104	&      &      & -1.92 & -1.78 & -1.85 & 0.87 & 0.68 & -0.19 & 	   & 0.15 & 0.46 & 0.74 & -0.53 \\
106	&      &      & -2.31 & -2.01 & -2.16 &      & 0.71 & ~0.01 & 	   & 0.71 & 0.48 &      &       \\	
\multicolumn{14}{c}{Group B}\\
4	&      &      & -1.60 & -1.67 & -1.63 &      & 0.78 & ~0.31 &      & 0.70 &      & 0.66 & -0.23 \\
7	&      &      & -1.69 & -1.76 & -1.73 & 0.98 & 0.46 & ~0.30 & 0.66 &-0.06 & 0.90 & 0.54 & -0.43 \\
8	&      &      & -1.44 & -1.66 & -1.55 & 1.14 & 0.35 & ~0.37 &      & 0.71 & 0.57 & 0.64 &  0.08 \\
10	&      &      & -2.06 & -1.72 & -1.89 & 1.19 & 0.84 & -0.14 & 	   & 0.76 & 0.43 & 0.08 & -0.28 \\
14	& 0.27 & 0.08 & -2.38 & -1.69 & -2.03 & 1.40 & 0.39 & ~0.49 & 	   & 0.73 & 0.86 &      &       \\	
16	&      &      & -1.91 & -1.96 & -1.93 & 1.23 & 0.47 & ~0.29 & 	   &      & 0.60 & 0.43 & -0.41 \\
23	&      &      & -1.66 & -2.04 & -1.85 & 0.98 & 0.58 & ~0.22 & 0.88 & 0.18 & 0.49 & 0.36 & -0.35 \\
25	&      &      & -1.91 & -1.86 & -1.88 & 1.07 & 0.48 & -0.01 & 	   & 0.52 & 0.29 & 0.44 & -0.58 \\
29	&      &      & -1.96 & -1.80 & -1.88 & 1.13 & 0.73 & ~0.03 &      & 0.49 & 0.66 & 0.31 & -0.82 \\
35	&      &      & -1.95 & -2.05 & -2.00 & 1.19 & 0.54 & ~0.25 & 0.73 & 0.14 & 0.38 & 0.29 & -0.63 \\
46	&      &      & -1.73 & -1.75 & -1.74 & 1.01 & 0.64 & ~0.23 & 	   & 0.53 & 0.76 & 0.43 & -0.70 \\
52	&      &      & -1.79 & -1.70 & -1.74 &      & 0.82 & ~0.15 & 	   & 0.78 & 0.52 & 0.44 &       \\	
72	& 0.26 & 0.10 & -1.83 &       & -1.83 &      & 0.46 & ~0.02 &      & 0.67 &      & 0.67 & -0.46 \\
83	& 0.43 & 0.06 & -1.96 &       & -1.96 & 1.16 & 0.54 & ~0.24 & 	   & 0.76 & 0.53 &      &       \\	
84	&      &      & -2.13 & -1.79 & -1.96 &      & 0.49 & ~0.19 &      & 0.61 & 0.44 & 0.40 & -0.95 \\
88	& 0.44 & 0.06 & -1.85 & -1.86 & -1.86 & 1.22 & 0.19 & ~0.15 & 	   & 0.54 & 0.56 & 0.24 & -0.41 \\
96	&      &      & -2.28 & -1.60 & -1.94 & 0.99 & 0.40 & ~0.03 & 	   &      & 0.52 &      & -0.49 \\
108	&      &      & -1.86 & -1.65 & -1.76 & 1.01 & 0.69 & ~0.23 & 	   & 0.79 & 1.32 & 0.51 & -0.26 \\
107	& 0.28 & 0.08 & -1.71 &       & -1.71 & 1.08 & 0.41 & ~0.42 & 	   & 0.56 & 0.61 & 0.57 &       \\	
111	&      &      & -2.05 & -1.80 & -1.92 & 1.16 & 0.68 & -0.03 & 	   & 0.67 &      & 0.75 & -0.87 \\
117	& 0.39 & 0.07 & -2.05 & -1.65 & -1.85 & 0.92 & 0.53 & ~0.30 & 	   & 0.84 & 0.45 & 0.63 &       \\	
118	& 0.41 & 0.07 &       &       &       & 1.23 & 0.59 & ~0.30 & 	   & 0.31 & 0.73 &      &       \\	
124	& 0.42 & 0.05 & -1.45 &       & -1.45 & 1.05 & 0.44 & ~0.34 & 	   & 0.05 & 0.48 &      & -0.67 \\
\multicolumn{14}{c}{Group C}\\
5	& 0.41 & 0.06 & -1.76 & -1.56 & -1.66 & 1.09 & 0.24 & ~0.31 & 	   & 0.60 & 0.20 &      & -0.41 \\
47	& 0.36 & 0.07 & -1.66 & -1.91 & -1.78 & 1.01 & 0.45 & ~0.38 & 	   & 0.54 & 0.27 &      & -0.03 \\
62	& 0.27 & 0.08 & -1.59 & -1.49 & -1.54 & 0.91 & 0.47 & ~0.34 & 	   &      & 0.00 & 0.45 & -0.75 \\
89	& 0.32 & 0.05 & -1.25 & -1.83 & -1.54 & 1.06 & 0.37 & ~0.64 & 	   &      & 0.42 &      &       \\	
94	& 0.28 & 0.08 & -1.64 &       & -1.64 & 0.97 & 0.83 & ~0.12 & 	   & 0.31 & 0.69 & 0.39 &  0.06 \\
95	& 0.41 & 0.05 & -1.52 & -1.84 & -1.68 & 1.11 & 0.57 & ~0.47 & 	   & 0.60 & 0.31 &      &  0.21 \\
100	& 0.23 & 0.06 & -1.65 & -1.86 & -1.76 & 1.10 & 0.36 & ~0.21 & 	   & 0.47 & 0.14 &      &  0.12 \\
105	& 0.34 & 0.06 & -1.32 & -1.70 & -1.51 & 1.07 & 0.70 & ~0.56 & 	   & 0.10 & 0.32 & 0.60 &       \\	
110	& 0.44 & 0.05 & -2.11 & -1.64 & -1.87 & 0.95 & 0.78 & ~0.17 & 	   & 0.11 & 0.51 &      & -0.48 \\
112	& 0.30 & 0.06 & -1.27 & -1.91 & -1.59 & 1.12 & 0.99 & ~0.52 & 	   & 0.88 & 0.63 & 0.44 &       \\	
113	& 0.43 & 0.05 & -1.72 & -1.63 & -1.68 & 1.09 & 0.20 & ~0.45 &      &-0.27 & 0.44 & 0.38 &       \\	
115	& 0.21 & 0.05 &       &       &       & 1.19 & 0.47 & ~0.61 & 	   & 0.27 & 0.37 &      &       \\	
116	& 0.20 & 0.04 & -1.23 & -1.57 & -1.40 & 1.17 & 0.28 & ~0.63 & 	   &-0.06 &-0.15 &      &       \\	
121	& 0.26 & 0.06 & -1.29 & -1.56 & -1.43 & 1.01 & 0.80 & ~0.50 & 	   & 0.71 & 0.29 &      &       \\	
122	& 0.29 & 0.09 & -1.77 &       & -1.77 &      &      & ~0.20 & 	   &      & 0.39 &      &       \\	
123	& 0.25 & 0.06 & -1.70 & -1.46 & -1.58 & 1.08 & 0.57 & ~0.18 & 	   &      & 0.13 &      &       \\	
126	& 0.44 & 0.05 & -1.76 & -1.78 & -1.77 & 1.08 & 0.86 & ~0.20 &      & 0.23 & 0.29 &      &       \\	
127	& 0.37 & 0.06 & -1.65 &       & -1.65 & 1.15 & 0.52 & ~0.48 & 	   &-0.02 & 0.40 & 0.70 &       \\	
128	& 0.30 & 0.06 & -1.64 &       & -1.64 & 1.08 & 0.38 & ~0.55 & 	   & 0.31 & 0.50 &-0.08 &       \\	
129	& 0.36 & 0.05 & -1.33 & -1.59 & -1.46 & 0.99 & 0.33 & ~0.51 & 	   & 0.55 & 0.20 &      &       \\	
\hline
\end{tabular}
\end{scriptsize}
\label{t:tab4}
\end{table*} 

\begin{center}
\begin{figure}
\includegraphics[width=8.8cm]{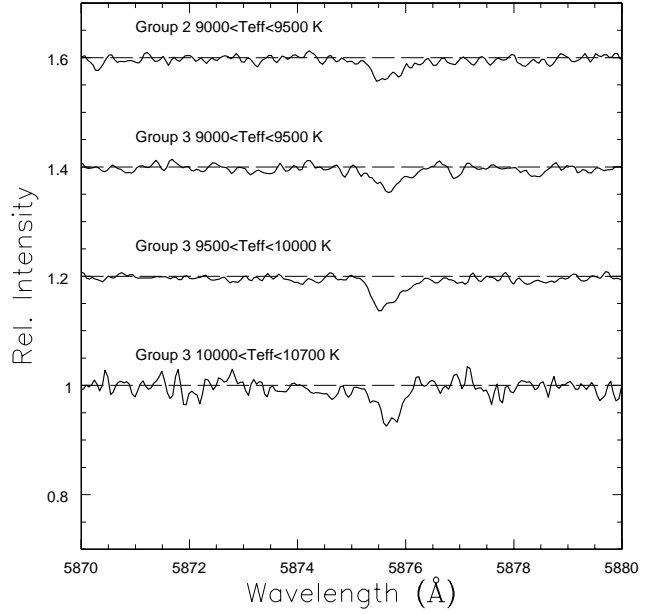}
\caption{Average spectra in the region of the He~I line at 5876~\AA\ for stars in different
temperature bins. See Section 4 for the definition of different groups of stars. Spectra have been offset
for clarity.}
\label{f:helines}
\end{figure}
\end{center}

\begin{center}
\begin{figure}
\includegraphics[width=8.8cm]{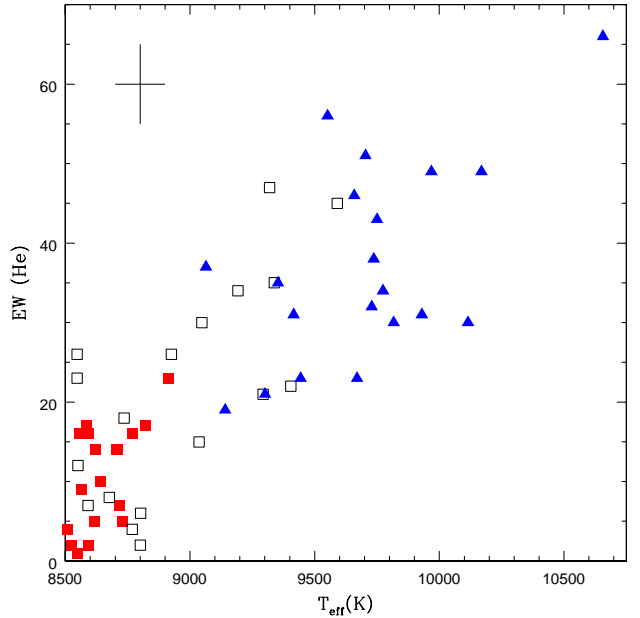}
\caption{\teff vs the equivalent width of the He I line at 5876~\AA. Different colours are for stars of
different groups (see Section 4): Group 1: 
red filled squares; Group 2: black open squares; Group 3: blue filled triangles. Typical error bars are also shown.}
\label{f:fig5}
\end{figure}
\end{center}

\begin{center}
\begin{figure}
\includegraphics[width=8.8cm]{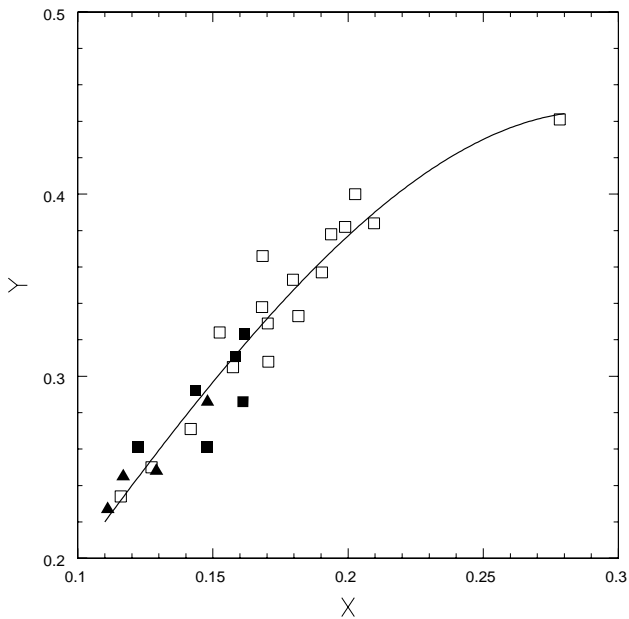}
\caption{Relation between the $X$\ index for the strength of the He~I line at 5876~\AA\ (see Eq. (2) for
a definition) and the He abundance by mass $Y$\ for BHB stars in NGC~2808 (Marino et al. 2013b: open squares), 
NGC~6752 (Villanova et al. 2009: filled triangles), and M~4 (Villanova et al. 2012: filled squares). He 
abundances from these two last papers were corrected for departures from LTE and a zero point offset of 0.036 
in $Y$. Overimposed is the best fit cubic through the origin (Eq. (3)).}
\label{f:fighe}
\end{figure}
\end{center}

\begin{table*}[htb]
\centering
\caption[]{Average He abundances from the 5876~\AA\ line for selected clusters}
\setlength{\tabcolsep}{1.5mm}
\begin{tabular}{lcccccccccc}
\hline
Cluster &$[$Fe/H$]$&Rel. Age (1) &
$\log{T_{\rm eff}}$(HB)&$\log{T_{\rm eff}}$(HB)&$\log{T_{\rm eff}}$(HB)&
Ref. & N$_{\rm stars}$ & $\log{T_{\rm eff}}$&  $<Y>$& r.m.s. \\
&&& Min & Median & Max &&& Range \\
\hline
NGC1851 & -1.18 &0.81 & 3.73 & 3.74 & 4.08 & 2 & 19 & 3.95$\div$4.06 &$0.297\pm 0.020$ & 0.088  \\ 
NGC2808 & -1.18 &0.83 & 3.75 & 3.92 & 4.57 & 3 & 17 & 3.96$\div$4.06 &$0.336\pm 0.013$ & 0.052  \\ 
M~5     & -1.33 &0.85 & 3.76 & 3.89 & 4.18 & 4 & 15 & 3.95$\div$4.02 &$0.312\pm 0.017$ & 0.064  \\ 
M~4     & -1.18 &0.97 & 3.72 & 3.76 & 4.04 & 5 &  6 & 3.95$\div$3.98 &$0.295\pm 0.011$ & 0.028  \\ 
M~22    & -1.70 &1.06 & 3.82 & 3.97 & 4.22 & 6 & 29 & 3.95$\div$4.03 &$0.338\pm 0.014$ & 0.074  \\ 
NGC6752 & -1.55 &1.02 & 3.82 & 4.02 & 4.47 & 7 &  4 & 3.93$\div$3.94 &$0.252\pm 0.016$ & 0.031  \\ 
\hline
\end{tabular}
\\
1. From Gratton et al. (2010); 
2. Gratton et al. (2012a); 
3. Marino et al. (2013b);
4. Gratton et al. (2013);
5. Villanova et al. (2012);
6. This paper;
7. Villanova et al. (2009)
\label{t:tabhe}
\end{table*} 

\subsection{Helium abundances}

Following Villanova et al. (2009), we derived He abundances for stars with an effective
temperature in the range $9000<T_{\rm eff}<11000$~K from the He I line (actually a 
narrow multiplet) at 5875.6~\AA). Figure~\ref{f:helines} shows some examples of the He
lines. To show them more clearly, we averaged spectra of different stars in bins in temperature.
Figure~\ref{f:fig5} shows the run of the $EW$\ of this line with temperature
for individual stars. Given the rather low S/N ($\sim 50$) of the spectra and the 
weakness of the He line, He abundances for individual stars have quite large errors. 

Marino et al. (2013b) present a non-LTE analysis of the He lines in BHB stars of 
NGC~2808. They also used a different code to synthesize He lines. The average He 
they obtained for their stars in NGC~2808 is quite high, $Y=0.34\pm 0.01\pm 0.05$, 
where the first error bar is derived from star-to-star scatter, and the second one 
describes the effects of systematics. Marino et al. determined an He abundance that 
is consistent with the one used to compute the stellar atmosphere, while in our previous 
analysis we assumed He to be a tracer element; that is, the model atmosphere is 
assumed independently of the He abundance that is derived. Our approach may lead to 
(unphysical) very large He abundances when the strength of the He line is overestimated 
due to measuring errors. The error bar obtained using Marino et al. approach is then
by far more realistic than obtained assuming it is a trace element. Furthermore,
we assumed LTE, while Marino et al. compute full statistical equilibrium
calculations. It appears that their methods are superior to 
those that we used in previous papers of this series, so it is interesting to obtain 
abundances on their scale.

We do not have access to their analysis code. However, we expect strong regularities
in the He abundances when they are derived from the same line in a limited range of
parameters, so we proceeded as follows. First, we examined the run of the non-LTE
corrections. We found that these are closely related to the $EW$
of the 5876~\AA\ He line and are represented well by a simple linear relation:
\begin{equation}
Y({\rm non-LTE}) - Y({\rm LTE}) = -9.00\times 10^{-4}~EW - 0.0192,
\end{equation}
where the $EW$\ is in m\AA. We could correct the helium LTE abundances for NGC6752 (Villanova 
et al. 2009), M4 (Villanova et al. 2012) using this 
formula. When added to the results by Marino et al. (2013b) for NGC~2808, we thus have
a consistent set of He non-LTE abundances for stars over the whole range of temperatures 
considered. 

The He abundances for BHB stars obtained in this way are mainly a function of $EW$\ and 
temperature, with a small correction for surface gravity and an even smaller one for metal 
abundance. To show this, we constructed a parameter $X$\ that is a combination of $EW$\ and
temperatures:
\begin{equation}
X = 10^{12}~EW/(T_{\rm eff}-5000)^{3.9},
\end{equation}
and then plotted the values of the He abundance by mass $Y$\ against $X$ (see Figure~\ref{f:fighe}).
We found that the points display a very small scatter around a cubic fit through the origin
\begin{equation}
Y=1.8224~X+3.2125~X^2-14.484~X^3,
\end{equation}
derived from stars cooler than the Grundahl jump. The r.m.s. of the points around this 
relation is 0.016 in $Y$. 

We derived He abundances for M~22 stars using these relations (Column 2 of Table~\ref{t:tab4}, 
with errors on the next column), and found an average He abundances by mass of $Y=0.338\pm 
0.014\pm 0.05$, where the first error bar is derived from star-to-star scatter, and the second 
one referring to systematics is simply the one adopted by Marino et al. (2013b) for 
NGC~2808 stars. This value is greater than expected from primordial nucleosynthesis ($Y=0.248$:
Cyburt 2004), even if the effect of first dredge-up is taken into account (the
expected surface He enhancement is $\Delta Y\sim 0.015$: Sweigart 1987). This indicates that
the BHB stars of M~22 hotter than 9000~K are moderately He-rich.

We notice that the He abundances of the BHB stars of M~22 are not significantly 
different from those of the BHB stars in the same temperature range in NGC~2808. This 
result could also be derived immediately by a comparison of the equivalent widths, which 
on average are similar at a given \teff.

To verify that the high He abundances found for the BHB stars of M~22
are not simply an artefact of our procedure, we used the same approach to homogeneously 
determine He abundances from the $EW$ of the 5876~\AA\ line for BHB stars in different 
clusters (see Table~\ref{t:tabhe}). The He abundances for NGC~1851 and 
M~5 we give here were derived using the formula given in this paper, and are then different from those
given in our previous papers. To put these He abundances into a context, we also 
listed values for the metallicity and relative age of each cluster as listed by Gratton 
et al. (2010). We also transformed the maximum, median, and minimum colours of HB
stars from the same source into minimum, median, and maximum temperature along the HB, and 
compared these values with the range of temperature of the stars observed in various 
clusters. These values are listed in Columns 4, 5, and 6 of the Table. We also listed
the temperature range for the stars for which He abundances were derived. When looking 
at the He abundances listed in this table, it should be recalled that they do not 
refer to the whole cluster, but only to those HB stars that happen to be in the right 
temperature range ($8500<T_{\rm eff}<11500$~K). Typically, there are stars hotter 
(and then, possibly more He-rich) and cooler (more He-poor) than the examined stars. 
Namely, in NGC~1851 and M~4, the stars examined by Gratton et al. (2012a) and
Villanova et al. (2012) are among the hottest (and then probably He-rich) in the 
cluster, while in NGC~6752 they are among the coolest (and then probably He-normal) 
ones. In all other cases, the HB extends on both sides of the temperature range over 
which He abundances were determined. A correct interpretation of the result then needs 
a more detailed modelling for each cluster and consideration of the impact of the first 
dredge-up. 

However, a look at this table indicates that when the observed stars are the coolest
along the HB (the case of NGC~6752), we indeed recover an He abundance that is consistent
within the errors with the cosmological value. As a result, systematic errors should not be large.
On the other hand, there is a wide range of He abundances. In most cases, moderate He
excesses with respect to the cosmological values are obtained and are consistent with
the location of stars along the HB (see e.g. the discussion by Gratton et al. 2012a and
Joo \& Lee 2013 for NGC~1851). The values obtained for the stars in M~22 fall at the high 
extreme of this range, with an average similar to that of the stars of (similar temperature) 
in NGC~2808. It also agrees well with the value proposed by Joo \& Lee (2013) in order to
explain colours of HB stars. 

We conclude that while systematics errors are possibly not 
negligible, they should not hamper the conclusion that the BHB stars of M~22 with 
$T_{\rm eff}>9000$~K are He-rich. We come back on this issue in Section 5.1.
We notice that the He abundance we obtained for NGC~1851 is lower than what is
obtained for similar stars in NGC~2808, even though the clusters have similar ages
and chemical compositions. The difference is significant at about 2~$\sigma$\ level even; while
this difference might perhaps be attributed to some other difference between stars in these clusters
(e.g. different CNO/Fe ratios), we think it needs to be confirmed by more data before any 
strong conclusion can be drawn.

\begin{table}[htb]
\centering
\caption[]{Parameter for lines measured on HR03 set up}
\begin{tabular}{lccc}
\hline
Element & Wavelength & E.P. & $\log{gf}$ \\
        & (\AA)   & (eV) & \\
\hline
Mg~I  & 4057.52 & 4.34 & -0.90 \\  
Mg~I  & 4167.22 & 4.34 & -0.75 \\  
Si~II & 4128.06 & 9.83 & ~0.36 \\  
Si~II & 4130.90 & 9.84 & ~0.55 \\  
Ti~II & 4053.84 & 1.88 & -1.13 \\  
Ti~II & 4163.61 & 2.58 & -0.13 \\  
Ti~II & 4171.86 & 2.59 & -0.29 \\  
Fe~I  & 4046.11 & 1.49 & ~0.28 \\
Fe~I  & 4063.60 & 1.56 & ~0.06 \\
Fe~I  & 4071.94 & 1.61 & -0.02 \\
Fe~I  & 4132.03 & 1.61 & -0.68 \\
Fe~I  & 4143.88 & 1.56 & -0.51 \\
Fe~I  & 4201.90 & 1.49 & -0.71 \\
Fe~II & 4173.44 & 2.58 & -2.16 \\  
Fe~II & 4178.86 & 2.58 & -2.44 \\  
Sr~II & 4077.91 & 0.00 & ~0.15 \\  
\hline
\end{tabular}
\label{t:tab5}
\end{table} 

\subsection{Metal abundances}

Abundances for other elements are given in Columns 6-16 of Table~\ref{t:tab4}. As 
described in Gratton et al. (2011, 2012a, 2013), we obtained O abundances from the 
high-excitation O~I triplet at 7771-74~\AA and Na abundances mainly from the D resonance
doublet at 5890-96~\AA. For a few cool stars we could also detect the higher excitation
Na~I line at 8194~\AA\ line (after appropriate correction for telluric lines). Consistently 
with the previous papers, abundances from these lines included non-LTE corrections following 
Takeda (1997) and Mashonkina et al. (2000).

As we did for NGC~1851 (Gratton et al. 2012a) and M~5 (Gratton et al. 2013), N 
abundances were also obtained using the high-excitation lines at 8216 and 8242~\AA. 
Analysis of these lines also includes non-LTE corrections, following Przybilla \& Butler 
(2001; see Gratton et al. 2012a) and appropriate correction for the contaminating 
telluric lines. The HR19A set up also allowed Mg abundances to be derived
from the Mg~II lines at 7877 and 7896~\AA. Atomic parameters for all these lines were 
the same as used in the previous papers. 
Several more lines were detectable in the blue spectra provided by the HR03 set up (see
Table~\ref{t:tab5}); and their oscillator strengths were taken from the NIST database.\footnote{
http://physics.nist.gov/PhysRefData/ASD/lines$\_$form.html }

The use of the LTE approximation for the analysis of these elements may be questioned.
For instance, Marino et al. (2013a) present both LTE and non-LTE abundances for Fe in 
their analysis of cool BHB stars in M~22. The non-LTE corrections were very small for
Fe~II ($<0.04$~dex, non-LTE abundances being larger) and a bit larger for Fe~I
(in the 0.2-0.3~dex range, non-LTE abundances being lower). When they applied these
corrections, they found consistent abundances from Fe I and Fe II lines.

We found abundances from Fe~I lines to be very similar to those from Fe~II lines, at
variance with the results by Marino et al. (2013a).
This is not due to differences in temperatures and gravities, which are quite similar
in the two analyses; it might rather be due to our adopting much higher values for the 
microturbulent velocity, because the Fe~I lines we used in our analysis are typically 
stronger than the Fe~II lines. Since microturbulent velocities are not derived from
first principles, but simply modified in such a way as to obtain agreement between abundances
derived from lines of different strength, and since non-LTE corrections are expected to be
larger for stronger lines, it is difficult to separate the two effects. Practically speaking,
in this case an LTE analysis with a high microturbulence produces abundances similar to a 
non-LTE analysis with lower microturbulence. Regardless the reason, we find that applying 
non-LTE  corrections as large as those considered by Marino et al. (2013) would destroy 
the agreement we obtain between Fe~I and II abundances, so we prefer not to apply them.

Abundances from the Sr~II line at 4077~\AA\ do not include any correction for the increased 
opacity due to the wings of H$\delta$; however, we checked that such a correction is very small 
($\leq 0.01$~dex). Furthermore, no correction for departures from LTE was applied. The size 
and even the sign of these corrections are not clear. Dworetsky et al. (2008) suggest that 
they should be small, if any, for population I A-type stars. Similar results have been 
obtained from statistical equilibrium calculations (Mashonkina et al. 2007; Andrievsky et 
al. 2011; Bergemann et al. 2012; Hansen et al. 2013) for Sr~II lines in metal-poor stars, 
but these results are only available for $T_{\rm eff}<6400$~K, that is for stars much cooler 
than our programme stars, and small trends are present at the high-temperature, low-gravity 
extreme of the range of parameters explored in these papers. The abundances of Sr that we 
obtain are smaller by about 0.5~dex than those by Marino et al. (2012) for subgiants in M~22. 
The reason for this systematic offset is not clear. On one hand, we notice that the result by 
Marino et al. is obtained from spectra with moderate dispersion and should then be 
considered with some caution, their main focus being on the difference obtained for the two 
SGB branches rather than on the absolute values. On the other hand, we notice that For and 
Sneden (2010) have obtained a low [Sr/Fe] abundance ratio (on average [Sr/Fe]=-0.30) for field 
BHB stars from an LTE analysis similar to ours for M~22. The stars they considered have 
effective temperatures and metal abundances similar to those of the stars we are analysing in
M~22, and they adopted similar values for the microturbulent velocities. An even lower 
abundance of [Sr/Fe]=-0.7 was obtained by Ambika et al. (2004) for a supra-BHB star in M~13. 
This might suggest a trend toward underestimating Sr abundances in LTE analysis of low-gravity 
hot stars that might represent an extrapolation of the small trend observed in cooler stars 
by Andrievsky et al. (2013). Appropriate statistical equilibrium calculations are required to 
settle this point. However, the effect is not overwhelmingly strong, and we think that our LTE
Sr abundances can still be used to separate different groups of stars in M~22 and to 
internally compare production of elements through the various $n-$capture processes.

\begin{table}[htb]
\centering
\caption[]{Sensitivity of abundances on the atmospheric parameters and total errors}
\setlength{\tabcolsep}{1.5mm}
\begin{tabular}{lcccccc}
\hline
Element & \teff & $\log{g}$ & $v_t$ & $[$A/H$]$ & $EW$ & Total \\
        & (K) & & (km~s$^{-1}$) & & (m\AA) & \\
\hline
Error          & 100 & 0.05 & 1.0 & 0.2 & 5 &  \\
\hline
\multicolumn{7}{c}{Cool BHB star ($T_{\rm eff}\sim$8000~K)}\\
\hline
$[$Fe/H$]$~I   & ~0.084 & -0.006 & -0.070 & ~0.004 & ~0.060 & 0.13 \\ 
$[$Fe/H$]$~II  & ~0.029 & ~0.013 & -0.025 & ~0.004 & ~0.070 & 0.08 \\
$[$N/Fe$]$~I   & -0.068 & ~0.006 & ~0.038 & -0.004 & ~0.113 & 0.14 \\
$[$O/Fe$]$~I   & -0.006 & ~0.000 & -0.164 & -0.011 & ~0.064 & 0.18 \\
$[$Na/Fe$]$~I  & ~0.033 & -0.013 & -0.010 & ~0.001 & ~0.119 & 0.12 \\
$[$Mg/Fe$]$~I  & -0.001 & -0.010 & ~0.038 & ~0.003 & ~0.111 & 0.12 \\
$[$Mg/Fe$]$~II & -0.054 & ~0.002 & ~0.034 & -0.004 & ~0.153 & 0.17 \\
$[$Si/Fe$]$~II & -0.098 & ~0.010 & ~0.022 & -0.007 & ~0.044 & 0.11 \\
$[$Ti/Fe$]$~II & -0.025 & ~0.010 & ~0.026 & -0.000 & ~0.061 & 0.07 \\
$[$Sr/Fe$]$~II & ~0.030 & ~0.002 & ~0.000 & ~0.002 & ~0.143 & 0.15 \\
\hline
\multicolumn{7}{c}{Hot BHB star ($T_{\rm eff}\sim$10000~K)}\\
\hline
He $Y$(NLTE)   & -0.022 & ~0.013 & -0.011 & -0.011 & ~0.034 & 0.05 \\
$[$Fe/H$]$~I   & ~0.061 & -0.017 & -0.006 & ~0.009 & ~0.097 & 0.12 \\ 
$[$N/Fe$]$~I   & ~0.013 & ~0.017 & -0.010 & -0.006 & ~0.075 & 0.08 \\
$[$O/Fe$]$~I   & ~0.009 & ~0.018 & -0.123 & -0.012 & ~0.038 & 0.13 \\
$[$Na/Fe$]$~I  & ~0.044 & ~0.000 & -0.009 & -0.003 & ~0.062 & 0.08 \\
$[$Mg/Fe$]$~II & -0.016 & ~0.026 & -0.025 & -0.009 & ~0.100 & 0.11 \\
$[$Si/Fe$]$~II & -0.008 & ~0.035 & -0.034 & -0.012 & ~0.089 & 0.10 \\
\hline
\end{tabular}
\label{t:tab6}
\end{table} 

\subsection{Sensitivity of abundances on the atmospheric parameters}

The sensitivity of abundances on the adopted values for the atmospheric parameters is given 
in Table~\ref{t:tab6}. It was obtained as usual by changing each parameter separately and
repeating analysis of the abundances. We also considered the contribution to the error
due to uncertainties in the equivalent widths, divided by the square root of the typical
number of lines used in the analysis. The values were computed for typical uncertainties 
in each parameter, as determined in Section 3.1. Results are given for two stars 
(\#1 and \#129) at the extremes of the observed range of temperatures. 

Results for He are for the abundance by mass $Y$\ and for the simulated non-LTE analysis
of Section 3.3. For the other 
elements, typical uncertainties in the abundances are $\pm 0.1-0.2$~dex. In most cases, 
equivalent widths contribute significantly to final errors. Fe abundances are also affected 
by errors in the effective temperatures. Since abundances of N~I, Mg~II, and Si~II are not 
influenced much by temperature for the cooler HB stars, the ratios to Fe abundances have an 
opposite temperature dependence. Oxygen and, in less measure, Fe abundances are also sensitive 
to the adopted value for the microturbulent velocity.

\begin{table}[htb]
\centering
\caption[]{Average parameters for the three groups}
\setlength{\tabcolsep}{1.5mm}
\begin{tabular}{lcccccc}
\hline
Parameter & Group 1 & r.m.s. & Group 2 & r.m.s. & Group 3 & r.m.s. \\
\hline
\teff (K) & 8468  & 237  & 8860  & 367  & 9697  & 372  \\
\hline
$Y$               &   ..  &  ..  & 0.363 & 0.076 & 0.328 & 0.072 \\
\hline
$<[$Fe/H$]>$      & -1.87 & 0.13 & -1.83 & 0.15 & -1.63 & 0.13 \\ 
$[$Sr/Fe$]$~II    & -0.54 & 0.26 & -0.54 & 0.26 & -0.18 & 0.36 \\
$[$N/Fe$]$~I      & ~0.68 & 0.15 & ~1.11 & 0.12 & ~1.06 & 0.07 \\
$[$O/Fe$]$~I      & ~0.65 & 0.12 & ~0.54 & 0.16 & ~0.54 & 0.23 \\
$[$Na/Fe$]$~I     & -0.08 & 0.11 & ~0.20 & 0.16 & ~0.40 & 0.17 \\
\hline
$[$Mg/Fe$]$~I     &  0.73 & 0.19 &  0.72 & 0.11 &       &      \\
$[$Mg/Fe$]$~II    &  0.48 & 0.29 &  0.59 & 0.26 &  0.27 & 0.31 \\
$[$Si/Fe$]$~II    &  0.48 & 0.19 &  0.56 & 0.23 &  0.32 & 0.20 \\
$[$Ti/Fe$]$~II    &  0.35 & 0.15 &  0.43 & 0.17 &  0.51 & 0.15 \\
\hline
\end{tabular}
\label{t:tab7}
\end{table} 

\begin{center}
\begin{figure}
\includegraphics[width=8.8cm]{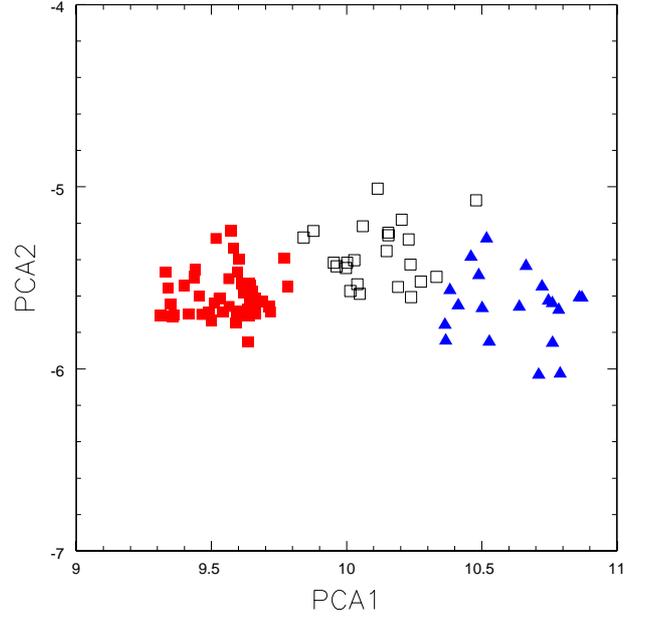}
\caption{Subdivision of the stars among the three groups in the two principal components plane.
Group 1 stars are represented by red filled squares, group 2 stars by open 
black squares, and group 3 stars by blue filled triangles. }
\label{f:figpca}
\end{figure}
\end{center}


\begin{center}
\begin{figure}
\includegraphics[width=8.8cm]{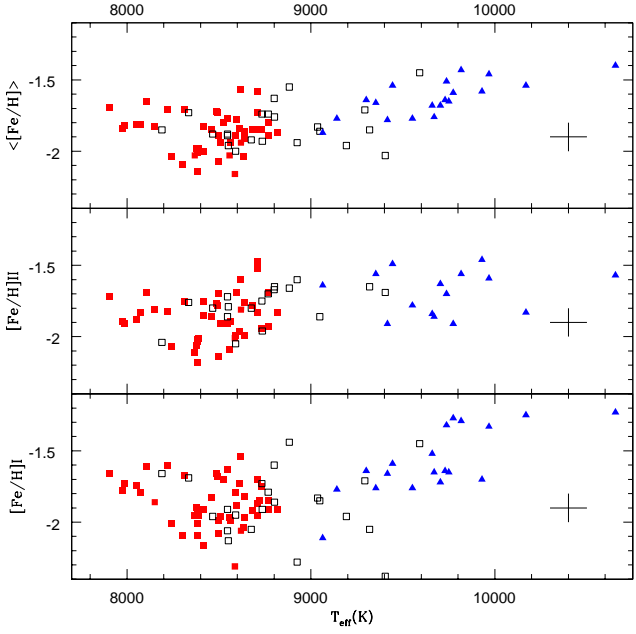}
\caption{\teff vs the abundances from Fe~I (lower panel), Fe~II lines 
(upper left panel, and the average of the two values (upper right panel). Different 
colours are for stars of different groups (see Section 4). Group 1: 
red filled squares; Group 2: black open squares; Group 3: blue filled triangles. 
Typical error bars are also shown.}
\label{f:fig6}
\end{figure}
\end{center}

\begin{center}
\begin{figure}
\includegraphics[width=8.8cm]{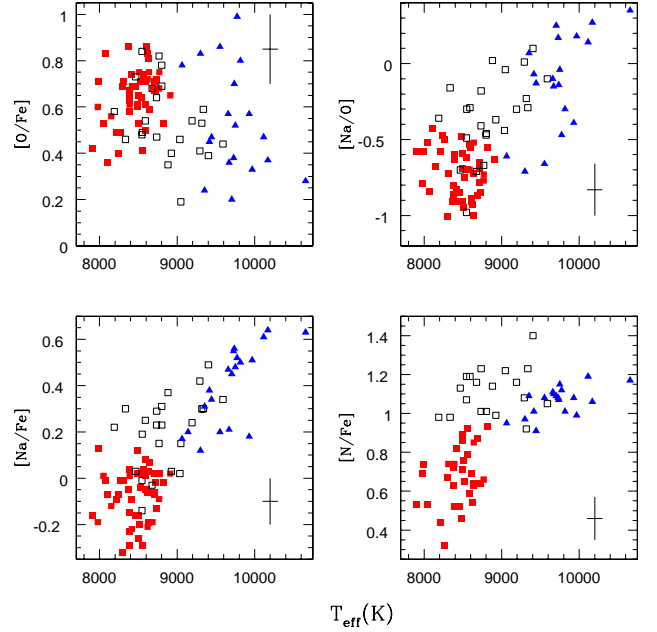}
\caption{\teff vs the abundances of N (lower right panel), Na (lower left panel), 
and O (upper right panel) and the [Na/O] abundance ratio (upper left panel. Different colours are for stars of
different groups (see Section 4). Group 1: 
red filled squares; Group 2: black open squares; Group 3: blue filled triangles.
Typical error bars are also shown. }
\label{f:fig8}
\end{figure}
\end{center}

\begin{center}
\begin{figure}
\includegraphics[width=8.8cm]{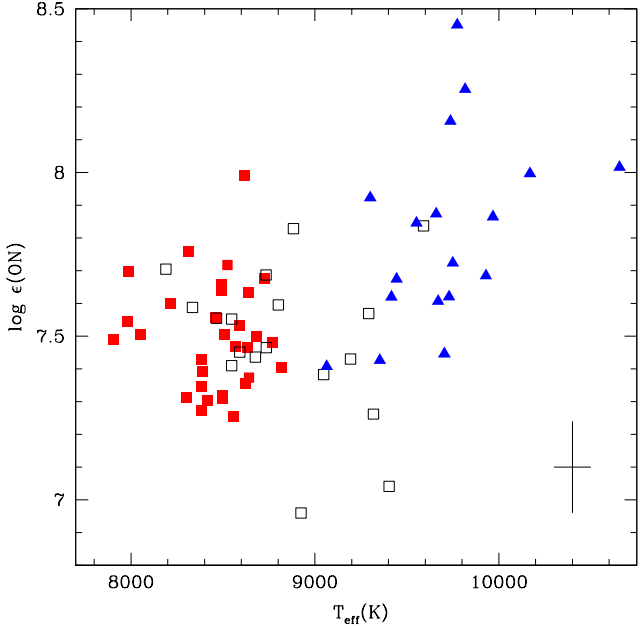}
\caption{\teff vs the sum of N and O abundances. Different colours are for stars of
different groups (see Section 4). Group 1: 
red filled squares; Group 2: black open squares; Group 3: blue filled triangles. }
\label{f:fig9}
\end{figure}
\end{center}

\begin{center}
\begin{figure}
\includegraphics[width=8.8cm]{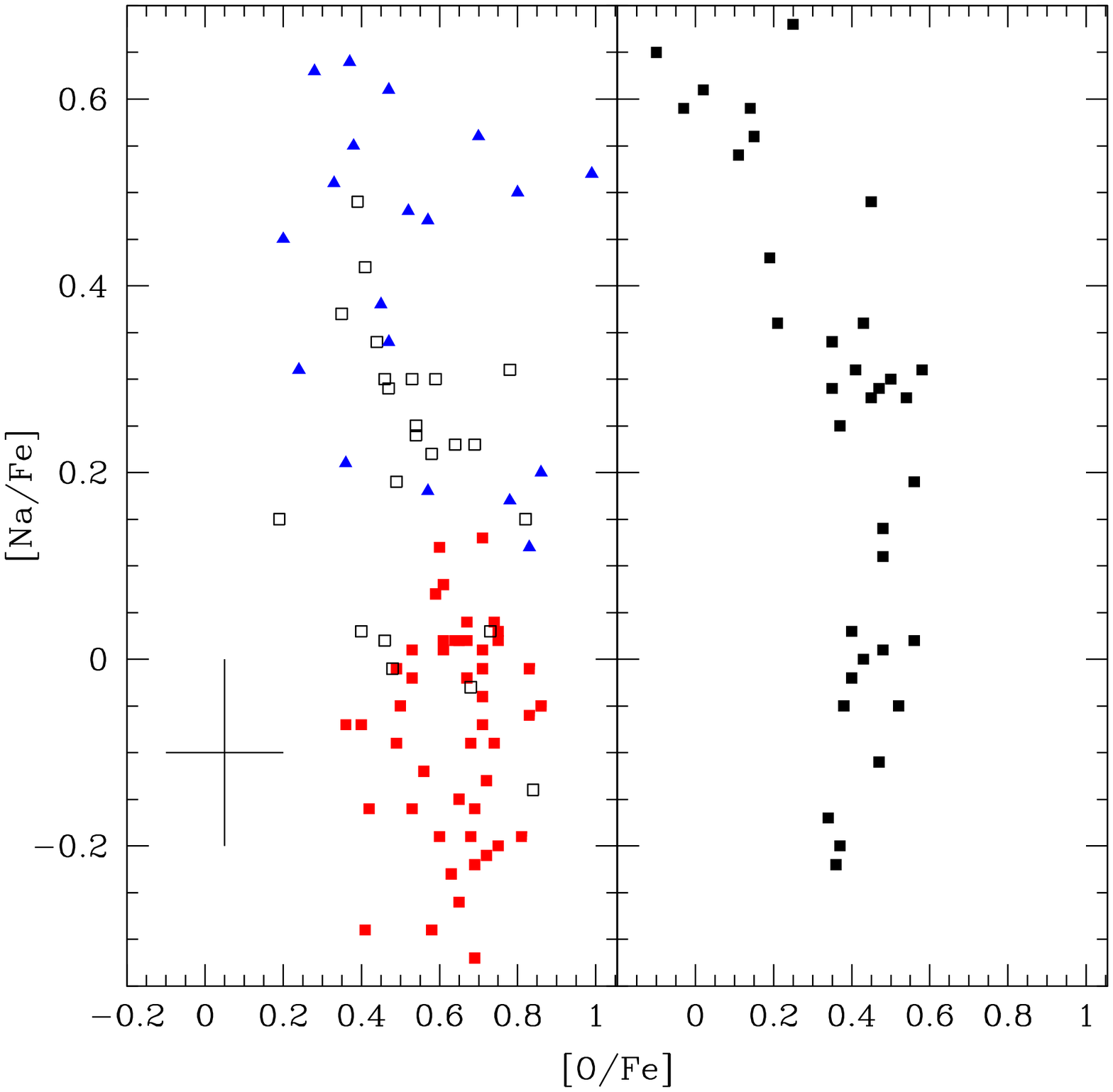}
\caption{[O/Fe] vs [Na/Fe] for BHB stars (left panel) and RGB stars from Marino et al. 
(2011a: right panel). In the left panel different colours are for stars of different 
groups (see Section 4). Group 1: red filled squares; Group 2: black open squares; 
Group 3: blue filled triangles. Typical error bars for our analysis are also shown.}
\label{f:fig10}
\end{figure}
\end{center}

\section{Cluster analysis and identifications of the main populations}

Previous work on M~22 has shown that there are several different populations in this
GC. Marino et al. (2009, 2011) find two main populations, each one
with a different metal content. These two populations can be well discerned along the
RGB and the SGB (see also Marino et al. 2012). Each one of these main populations
displays a spread over the Na-O anti-correlation, showing that they
have a fine structure (see Marino et al. 2011). Understanding the HB of M~22 requires 
identifying the progeny of these different populations during the He-core
burning phase. To this purpose we should first remember that we did not
observe all HB stars in our study. It is then probable that our sample does not represent
the whole cluster population.

Our approach was then to identify natural groups among the observed stars. This was
done using a statistical cluster analysis. We used the {\it k-means}
algorithm (Steinhaus 1956; MacQueen 1967) as implemented in the $R$\ statistical 
package ($R$\ Development Core Team 2011), where $R$\ is a system for statistical computation 
and graphics, freely available on-line\footnote{http://www.R-project.org}. The 
following parameters were considered when performing the analysis: effective temperatures 
(to avoid a variance that is too dissimilar to those of other quantities, 
we used $T_{\rm eff}/10000$~K in the analysis) and absolute visual magnitude $M_V$, 
which describe the location of stars along the HB; and four parameters describing the chemical
composition of the stars ($<[$Fe/H$]>$, [N/H], [O/H], [Na/H]). The
first parameters are related to the main subdivisions in
metal-poor and metal-rich populations, which are discernible from photometry
of the SGB and RGB (Marino et al. 2009, 2012), while the three remaining ones are
related to the Na-O anti-correlation.

We found that a subdivision in three groups is able to capture a large fraction of
the information, the variance between groups representing 66.7~\% of the total
variance. The three groups are made of 49, 23, and 20 stars. Figure~\ref{f:figpca}
shows the subdivision of the stars into the three groups on the plane of the two
principal components. A principal component analysis is the first step of a statistical
cluster analysis. The two principal components are defined as follows:
\begin{eqnarray}
PCA1 & = & 1.11~10^{-5}~T_{\rm eff}
      +0.151~M_V
      +0.209~\log{n({\rm Fe})} \nonumber \\
     && +0.068~\log{n({\rm O})}
      +0.656~\log{n({\rm Na})}
      +0.697~\log{n({\rm N})} \nonumber
\end{eqnarray}
and
\begin{eqnarray}
PCA2 & =&-4.70~10^{-6}~T_{\rm eff}
      -0.262~M_V
      -0.340~\log{n({\rm Fe})} \nonumber \\
&&      -0.738~\log{n({\rm O})}
      -0.238~\log{n({\rm Na})}
      +0.461~\log{n({\rm N})} \nonumber
\end{eqnarray}

The main parameters of the three groups are listed in Table~\ref{t:tab7}. The r.m.s.
for the abundances of the individual groups agree well with the scatter expected 
for uncertainties in the analysis listed in the last column of Table~\ref{t:tab6},
in agreement with expectations if these groups have a physical meaning. The main 
driver for the subdivision into these groups is the correlated variations in N and Na 
abundance, which are responsible for most of the variations in PCA1, and PCA2 is a 
combination of several parameters with more weight on N and O abundances. 
Figure~\ref{f:figpca} shows that the subdivision into groups is driven by PCA1.
This means that the subdivision in groups is mainly based on the chemistry.
Figures~\ref{f:fig6} and ~\ref{f:fig8} give the run 
of the abundances of Fe, Na, N, O with effective temperature, with different symbols 
for the three groups. The separation is quite evident. Figure~\ref{f:fig9} gives the same
run for the sum of N and O abundances 
with temperature. The subdivision into groups is still clear. 
We notice that effective temperature plays a minor role in the definition of PCA1. 
That there is quite a clear segregation of the groups in the 
colour-magnitude diagram is essentially a consequence of the fact that stars with 
different chemistry occupy different locations along the horizontal branch.
Finally, Figure~\ref{f:fig10} 
compares the Na-O anti-correlation found for the HB stars observed in this paper with 
the one obtained for the RGB by Marino et al. (2011a).

From this data, we found that Groups 1 and 2 are made of metal-poor 
stars. Abundances for these HB stars match the abundances found by 
Marino et al. (2011) very well for their metal-poor population, and this can be identified with 
the b-SGB population. We obtained low Sr abundances for these two groups, again 
in good agreement with the result by Marino et al. The two groups also differ for 
N and Na abundance, the first group being poorer in these elements than the second 
one. The low abundances of N and Na of the first group are compatible with the one for 
evolved metal-poor giants in the field (see e.g. Gratton et al. 2000). These stars 
can then be identified with the first (or primordial) population in the cluster.
This group is also slightly more O-rich than the second one; however, we did
not find any extreme O-poor stars in our sample, while some of them were found by
Marino et al. (2011b). We then think that the second group is made of stars
that have an intermediate position along the Na-O anti-correlation and that
stars strongly depleted in O are not sampled in our analysis because they fall in
the region of the HB hotter than the Grundahl jump.

Group 3 is made up of metal-rich stars. Again, the abundances match those found by 
Marino et al. (2011b) very well for this group of stars, which can be 
identified with the f-SGB population. We notice that this group is rich in Sr, as 
found by Marino et al. (2012) for the f-SGB stars of M~22 (although there is an 
offset between the Sr abundances we obtain for HB stars and those found by Marino 
et al. for SGB stars; see Section 3.4). This group of stars is rich in Na and N,
but it is also quite rich in O. We think they can be identified with the
less O-poor/Na-rich stars along the Na-O anti-correlation for the metal-rich
stars of Marino et al. (2011b). As for the metal-poor stars, Marino et al. (2011b)
also find evidence for a metal-rich population strongly depleted in O and very
rich in Na. Again, we think that this population is not sampled by our analysis
because these stars also fall in the HB-region that is hotter than the Grundahl jump.

We finally notice that stars of both Groups 2 and 3 seem overabundant in He.
The result for Group 2 is more uncertain, because it is based on only eight stars that,
being not very hot, have quite weak He lines (see Figure~\ref{f:helines}). The
result for Group 3 is more robust, since it is based on 20 stars that have
stronger He lines. However, in both cases the internal scatter is not much larger
than expected for the errors listed in Table~\ref{t:tab6}. We then suggest that
stars in both these groups are overabundant in He with respect to the cosmological
value.

\begin{table}[htb]
\centering
\caption[]{Number of stars of different populations of M~22: 
FG=first generation with primordial composition (normal He, low Na, high O); 
SG-I=second generation with intermediate composition (moderately high He and Na, moderately
low O); SG-E=second generation with extreme characteristics (very high He and Na, very
low O)}
\begin{tabular}{lcccc}
\hline
Population & FG & SG-I & SG-E & Total \\
\hline
\multicolumn{5}{c}{RGB: Marino et al. (2011a)}\\
\hline
Metal-poor & 12 &  ~4  &  ~5  &   21 \\
Metal-Rich &  - &  ~9  &  ~5  &   14 \\
\hline
\multicolumn{5}{c}{Frequency over the RGB and SGB (Marino et al. 2011a, 2011b)}\\
\hline
Metal-poor & 0.34 & 0.11  & 0.14  &   0.60 \\
Metal-rich &  -  &  0.26  & 0.14  &   0.40 \\
\hline
\multicolumn{5}{c}{BHB: This paper}\\
\hline
Metal-poor & 49 &  23  &   -  &   72 \\
Metal-rich &  - &  20  &   -  &   20 \\
\hline
\multicolumn{5}{c}{Frequency over the whole HB}\\
\hline
Metal-poor & 0.39 &   0.15 & $\sim$0.14  &  $\sim 0.68$ \\
Metal-rich &  -   &$>$0.14 & $\sim$0.14  &  $\sim 0.32$ \\
\hline
\end{tabular}
\label{t:tab8}
\end{table} 

\begin{center}
\begin{figure}
\includegraphics[width=8.8cm]{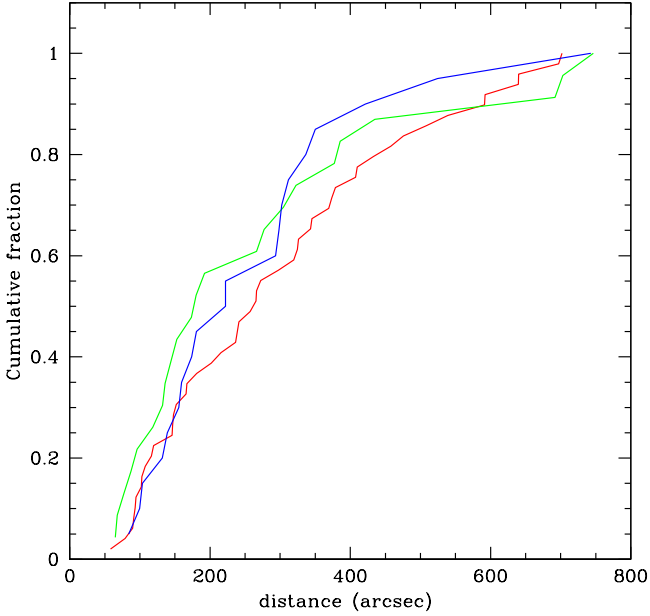}
\caption{Cumulative distribution of stars in different groups with separation
from cluster centre. Different colours are for stars of
different groups (see Section 4): Group 1: red; Group 2: green;
Group 3: blue}
\label{f:figcumdist}
\end{figure}
\end{center}

\begin{center}
\begin{figure}
\includegraphics[width=8.8cm]{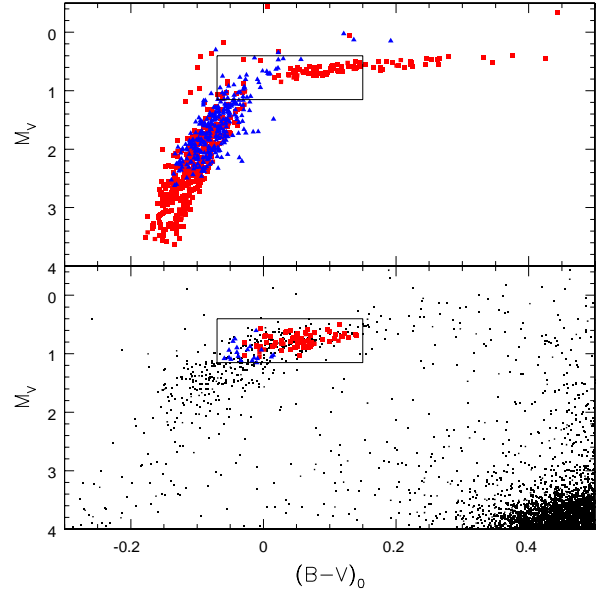}
\caption{Comparison between a synthetic (upper panel) and an observed (lower panel)
colour-magnitude diagram for the horizontal branch of M22. Red squares are
metal-poor stars (in the lower panel, stars of Groups 1 and 2); blue triangles
are metal-rich stars (in the lower panel, stars of Group 3). In the lower panel,
stars not observed in this paper are dots. The region of the HB observed in this
paper is marked with a rectangle in both panels}
\label{f:synt2}
\end{figure}
\end{center}

\begin{center}
\begin{figure}
\includegraphics[width=8.8cm]{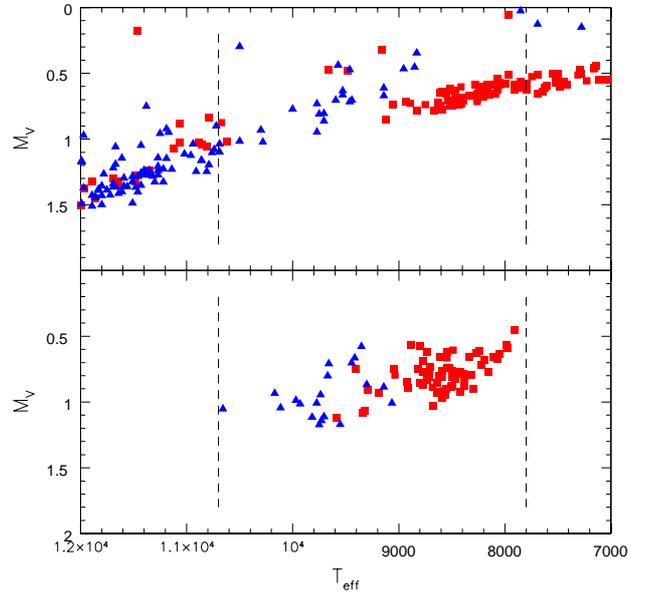}
\caption{Details of the comparison between synthetic (upper panel) and observed 
(lower panel) $T_{\rm eff}-M_V$\ diagram for the horizontal branch of M22. 
Red squares are metal-poor stars (in the lower panel, stars of groups 1 and 2); 
blue triangles are metal-rich stars (in the lower panel, stars of group 3).  
The region of the HB observed in this paper is within the dashed lines in both panels}
\label{f:synt}
\end{figure}
\end{center}

\section{Discussion}

We intend to make a quantitative comparison between the observed distribution of 
stars along the HB of M~22 and appropriate synthetic colour-magnitude diagrams, 
in order to shed light on the properties
of the different stellar populations of M~22. This comparison makes use of rough
estimates of the frequencies of these different populations that were derived as 
follows.

First, we recall that our observations are biased because we observe neither the
stars hotter than the Grundahl jump nor those in the RR Lyrae instability strip.
Owing to the possible presence of radial gradients, it would be important
to sample different cluster regions. In fact, Kunder et al. (2013)
find a clear indication that stars on the f-SGB are more centrally concentrated than 
those on the b-SGB in M~22. They did not find any clear trend for HB stars, but in 
that case their analysis was only limited to the annulus from 4 to 6~arcmin from the 
centre and a limited number of stars; for reference, the half-light radius of M~22 
is 3.36~arcmin according to Harris (1996), and gradients are expected to be
clearest when a wide range in logarithm of distance from centre is considered. For 
practical reasons (avoiding collisions of fibres and contamination by neighbours), 
most of our stars are at rather large distances from the cluster centre, within an 
annulus from 1 to 12.5~arcmin, with a median value of 4.0~arcmin. Figure~\ref{f:figcumdist} 
shows that Group 1 looks somewhat less concentrated than Groups 2 and 3. This would 
agree with expectations and with the result on the SGB by Kunder et al. (2013), because 
stars of the first generation are usually less concentrated than those of the second generation, 
and we identified Group 1 with the first generation. However, Kolmogorov-Smirnov tests
show that differences between groups are not significant, probably because we only
observed a few stars in each group. Since the effect is not overwhelming, we 
neglect radial variations in this discussion.

We then considered the statistics of different groups of stars from the analysis of 
Marino et al. (2011a). While there are only 35 stars in this sample, there are not 
strong evolutionary biases because stars with slightly different chemical compositions 
end up in similar locations along the RGB. They may then give a rough idea of the 
real frequencies of the main populations. Furthermore, as discussed in the previous 
section, abundances by Marino et al. (2011a) are on a scale that looks quite similar 
to what we obtain for the HB stars, so that a direct comparison is possible. 
Table~\ref{t:tab8} provides the results of these counts. In this table, FG are
O-rich, Na-poor stars that we attribute to the first generation of stars in M~22, and
SG-I and SG-E are second generation with intermediate or extreme values for Na and
O abundances throughout the Na-O anti-correlation (see Carretta et al. 2009). 

The FG stars are expected to have a He abundance close to the cosmological value; SG-I 
and SG-E are expected to be moderately and extremely He-rich stars. We then 
identified our Group 1 with the FG metal-poor population, Group 2 with the SG-I
metal-poor population, and Group 3 with the SG-I metal-rich population found on
the RGB using data by Marino et al. (2011a).

Using the photometry by Richter et al. (1999), which covers a region within 5~arcmin of 
the cluster centre with a median of 2.6~arcmin, we found that in their data there are 71 stars 
hotter than the Grundahl jump (that is, 32\% of the total HB stars), 16 stars (7\%) in 
the instability strip, and 138 (61\%) stars within the range of temperatures observed 
by us. These values are consistent with the star counts in the HB by Kunder et al. 
(2013).

If we make the assumption that stars in the instability strip are the cooler extension 
of our Group 1, in agreement with the properties of the RR Lyrae in M~22 (Kunder et al. 
2013), and are then FG metal-poor stars, we obtain that a fraction of 
$0.07+0.61\times (49/92)=0.39$\ of the HB stars of M~22 belongs to this population. Within the 
rather large errors due to small number statistics, this frequency agrees with the one 
found for this population along the RGB by Marino et al. (2011a). We also note that this 
group seems to coincide with the group that is redder than the gap at about 9000-9500~K noticed 
by Kunder et al. (2013) and that makes up 39\% of the HB stars of their sample.

We further assume that all SG-I metal-poor stars have temperatures within the range 
we observed. This is justified by the fact that we do not have any group 2 star close to 
edge of this range. In this case the frequency of metal-poor SG-I stars in the whole 
HB can be obtained by multiplying the frequency in our sample by the fraction of HB stars 
that are within the temperature range within our sample, that is, $0.61\times (23/92)=0.15$.

If we repeat a similar estimate for the SG-I metal-rich stars (identified with
our Group 3), we get a fraction of $0.61\times (20/92)=0.13$. However, this may be considered
more a lower limit than real data, because we may have missed stars of this
group because they are slightly hotter than the temperature limit of our survey. This is
also suggested by a comparison with the results by Kunder et al. (2013): our 
intermediate groups (including both Groups 2 and 3) seem in fact to coincide with the 
clump of stars at the HB position parameter $l_{\rm HB}\sim 26$\ in their analysis, which 
includes some 35-40\% of the HB stars of M~22. This is more than the fraction of
0.28 we obtain by summing the stars in our Groups 2 and 3. Also, there are stars in
this clump that are clearly hotter and fainter than the stars we observed.

On the other hand, we missed all SG-E stars that make up 28\% of the stars along the 
RGB observed by Marino et al. (2011a). They very likely end up on the HB with temperatures 
hotter than the Grundahl jump, which are 32\% of total according to the Str\"omgren 
photometry by Richter et al. (1999). According to Marino et al. (2011a), these hot
stars should be roughly equally divided among the metal-poor and the metal-rich populations. 

With these assumptions, we obtain the frequencies given in the last part of
Table~\ref{t:tab8}. These values should be considered with caution, because they
are based on low number statistics. However, on the whole they suggest that
some 60-70\% of the HB stars of M~22 belong to the metal-poor population and
30-40\% to the metal rich one. Given the large uncertainties, these values cannot
be considered to disagree with the overall 60-40 subdivision found by Marino et al.
(2011a) from RGB stars and a similar rough estimate for SGB stars by Marino et 
al. (2011b). We notice that a 70-30 subdivision has also been obtained recently
by Carretta et al. (2014, in preparation) who consider the bimodal distribution 
into two sequences along the RGB in the $(y, v-y)$\ colours from Richter et al. 
(1999) photometry, and by Joo \& Lee (2013) from a re-analysis of ground-based 
photometric data by Lee et al. (2009) and of HST-ACS data by Piotto (2009).

We note, in summary:
\begin{itemize}
\item All Group 1 stars (roughly 39\% of the cluster stars, with 7\% cooler than the blue edge of
the instability strip) are cooler than 8900~K. They should then have a very narrow mass
range, since all are more massive than 0.64~M$_\odot$. For the stars we observed (hotter
than the instability strip), the range in mass is from 0.66 to 0.64~M$_\odot$. They 
should have either all the same He abundance, essentially the cosmological value, or 
a very narrow range (less than 0.01 in $Y$).
\item Group 2 stars (roughly 15\% of the cluster stars) also have a limited range in 
temperature ($8300<T_{\rm eff}<9600$~K), which indicates a very limited range in 
mass too: $0.63<M<0.65$~M$_\odot$. They therefore are likely to all have nearly the same 
He abundances at $Y\sim 0.015$\ larger than Group 1 stars. This value 
is lower than indicated by the strength of the 5876~\AA\ line.
\item All Group 3 stars (roughly 18\% of the cluster stars) are hotter than 9000~K. Those 
stars that we observed are also cooler than 10700~K; however, there should be hotter 
stars in this group, up to $\sim 12000$~K. Depending on the CNO content, this sets 
an upper limit to the mass, but there should also be a narrow range around it, 
suggesting a unique value for He abundance. For normal [CNO/Fe], this means 
$0.60<M<0.62$~M$_\odot$ and $Y\sim 0.28$, and for CNO enhanced, this means a range in 
mass $0.57<M<0.59$~M$_\odot$ and $Y\sim 0.30$. Again, this value is lower than 
indicated by the strength of the 5876~\AA\ line, though the discrepancy is not as 
large as found for Group 2 stars if CNO is enhanced.
\item In addition, there should also be more He-rich stars in M~22, not observed by us
because hotter than 10700~K. They should make up about 28\% of the cluster. Using
photometry by Kunder et al. (2013), we found that half of them are hotter than 14000~K, 
that is, less massive than $\sim 0.54$~M$_\odot$. These stars should be very rich in He 
($Y>0.33$). The other half should have $10700<T_{\rm eff}<12000$~K, and a He abundance 
of $Y\sim 0.30$.
\end{itemize}

Based on the previous discussion, we prepared synthetic colour-magnitude diagrams
for the HB of M~22. They were performed as described in Salaris et al.~(2008) and
Dalessandro et al. (2011). We 
employed the HB evolutionary tracks for [Fe/H]=-1.8 as reference set for
the metal-poor stars, and [Fe/H]=-1.6 for the metal-rich ones from the BaSTI database
(Pietrinferni et al.~2006). In addition, we have interpolated among the 
$\alpha$-enhanced BaSTI models to determine HB tracks for various values of the 
helium content Y. Finally, we also considered both the reference set and the
CNONa anti-correlated models with CNO sum enhanced by 0.3~dex (Pietrinferni et al. 
2009). We adopted a distance modulus $(m-M)_V$=13.60 and E(B-V)=0.36 from Harris
(1996).

In our simulations, we have considered as constraints a number ratio
between bright (metal-poor) and faint (metal-rich) SGB stars equal to 60:40.
We notice that reproduction of the exact details of the colour/temperature distribution
requires a lot of fine tuning on the exact parameters. This fine tuning is very
time expensive and possibly misleading, because it might lead to over-confidence
on details that may instead depend on the way models were constructed. We then
focussed on the main features of such a comparison.
We made several attempts to fit both the HB and the other sequences in M~22
(RGB and SGB), changing mass loss and the range of He abundances for the various
populations. We find that continuous distributions in He cannot reproduce the
segregation between the three groups of stars observed along the HB of M~22.
Furthermore, the minimum He abundance of metal-rich stars should be greater than
the cosmological value, or else these stars would be fainter (at same
temperature) than metal-poor stars, which is not indicated by observations. We
then finally assumed the following recipe for the different populations of M~22:
\begin{itemize}
\item Metal poor: two subcomponents, the first one (corresponding to Group 1)
has a uniform distribution of He in the narrow range $0.246<Y<0.251$\ and a total
average mass lost along the RGB of 0.16~$M_\odot$, with a Gaussian distribution with an r.m.s.
of 0.01~$M_\odot$; and a second one (whose cooler part corresponds to Group 2)
has a uniform distribution of He in the narrow range $0.285<Y<0.319$\ and the same total
average mass loss of Group 1, but with a narrow Gaussian distribution with an r.m.s.
of 0.005~$M_\odot$
\item Metal rich: a uniform He distribution over the range $0.285<Y<0.325$\ and
a Gaussian distributed total mass loss with average value of 0.160~$M_\odot$\ and 
r.m.s. of 0.005~$M_\odot$
\end{itemize}
We assumed for both the metal-poor and metal-rich HB stellar components an RGB progenitor
whose age at the RGB tip is equal to 12~Gyr. Finally, a differential reddening of 0.05
mag (see Kunder et al. 2013) was added to the synthetic data.

We had to assume values of $Y$ that are actually lower than determined from our
spectroscopic data for the He-rich populations. In fact, if HB stars were even 
richer in He than assumed in the models, they would be brighter than observed. Also, 
we would expect them to be less massive and bluer than observed if they were coeval and lost the same 
amount of mass along the RGB than the He-normal stars. We conclude that while there is good
qualitative agreement between these different helium abundances, there is some
disagreement about the exact quantitative level, which is only partly surprising if
we consider that they are derived by using completely different and independent techniques.

Figures~\ref{f:synt2} and \ref{f:synt} compare the synthetic HBs with the observed
ones. We notice that this simulation reproduces (i) the ratios between stars within the 
RR Lyrae instability strip, blueward of it with $M_V<16$, and in the blue tail, i.e. 
$M_V>16$ measured by Kunder et al. (2013),
within Poisson errors; (ii) the number ratio between metal-poor and metal-rich stars
within the region considered in this paper; (iii) roughly, the segregation of the
stars of different groups in the $T_{\rm eff} - M_V$\ plane observed in this
paper; (iv) the average He abundance of metal-rich stars in the range of effective
temperature observed in this paper (the actual value of Y=0.29 is at the lower limit
of the range admitted by our data); and the number ratio of metal poor/metal rich stars
measured on the SGB by Marino et al. (2009).
The scatter in the observed diagram of Figure~\ref{f:synt2} looks larger than 
for theoretical colours. This is largely due to residual photometric errors
or some underestimate of the impact of differential reddening.
In fact, if we look at the distribution of stars with temperatures of Figure~\ref{f:synt}
(where the impact of photometric error and differential reddening is much
smaller), the agreement between observation and models is improved.

Answering a question from the referee, we noticed that the scatter in $M_V$\ 
around the mean $T_{\rm eff} - M_V$ relation ($\sim 0.10$~mag) is larger than predicted 
from evolutionary effects ($\sim 0.07$~mag) for the metal-poor stars, while it is actually 
smaller than predicted (0.15 vs 0.18~mag) for the metal-rich ones. The residual errors
in differential reddening ($\sim 0.01$~mag in E(B-V), that is, $\sim 0.05$~mag in
the residuals in $M_V$) may explain a part of the larger-than-expected scatter observed 
for metal-poor stars. Also, errors due to photometry should be considered. Finally, it is 
possible that the assumptions we made in our population synthesis represents an 
over-simplification of the real situation in M~22. However, we stress that our comparison 
only gives a sketch of the real properties of the populations in M~22.

On the whole, this interpretation of the HB of M~22 agrees with what is proposed by Joo
\& Lee (2013). Details are slightly different, because we find that there should be 
some overlap between the most He-rich stars of the metal-poor population and the 
moderately He-rich stars of the metal-rich one. Such fine detailing is only possible 
here because we have determined the chemical composition of individual stars, while 
Joo \& Lee could only work with photometric data. But excluding this detail, the 
agreement between ours and their description of the HB of M~22 is impressive.

\section{Conclusions}

We presented a spectroscopic abundance analysis of a sample of 92 blue HB stars
in M~22 in order to discuss the relation existing between chemical composition and the
location of the stars along the HB of globular clusters. The stars selected for the
analysis are in a restricted range of temperatures between 7800 and 10700~K. Cooler
stars were not considered because they are RR Lyrae variables, while surface abundances
for hotter stars are known to be heavily affected by microscopic diffusion and
radiative levitation. However, stars in our sample are representative of the majority
of HB stars in this clusters. We obtained spectra in three spectral ranges, including
the stronger lines of Na and O and of the $n-$capture element Sr. In addition, we were able to derive
abundances for He, N, and Fe (from both neutral and singly ionized lines), as well as for
other species (Mg, Si, and Ti). Whenever possible, reddening free effective temperatures 
were obtained from a calibration of the strength of H$\delta$; otherwise they were obtained
from visual and violet colours. Abundances of Na, O, N, and He include non-LTE corrections
obtained from literature calibrations. We did not apply non-LTE corrections to Fe
abundances because we get agreement between Fe I and Fe II abundances when assuming LTE.
This might be a consequence of the rather high values we adopted for the microturbulent
velocity, which do, however, agree with determinations for field BHB stars.
We get a rather high value for the He abundance, similar to the one recently obtained for
similar stars in NGC~2808 by Marino et al. (2013b), but higher than in other GCs for
which a similar analysis was performed.

We then applied a statistical cluster analysis to our data and found that the stars we
studied divide into three groups that occupy adjacent location along the HB, with some
overlap. The coolest group is metal-poor, Sr-poor, N- and Na-poor, and O-rich. This result 
confirms an earlier finding for a few stars by Marino et al. (2013a). The intermediate group 
is still metal-poor and Sr-poor, but is N- and Na-rich, and moderately O-poor. The hotter 
group is metal-rich and Sr-rich, moderately N- and Na-rich, but also O-rich. These three 
groups have a clear correspondence with the different populations found on the RGB and SGB 
by Marino et al. (2011b, 2012): the metal-poor and $s-$poor population that is also found along 
the RGB that is the progeny of the b-SGB, and the metal-rich and $s-$rich RGB one that is the 
progeny of the f-SGB. We do not find any extremely O-poor star in our sample, but HB stars
with this composition are expected to be hotter than the range we observed. Our result then
nicely confirms and extends previous investigations and supports the assumption that the spread
in colour of HB stars within a GC is mainly determined by variations in the chemical
composition, as measured by proxies like Na and O and whenever possible, directly by He
lines. We also found that there is not only qualitative agreement between predictions of this
scenario and observations, but also a quantitative one; furthermore, star counts in different
evolutionary phases agree with each other, supporting the proposed relation between different
groups of HB, RGB, and SGB stars.

We found that there should be fairly He-rich stars in M~22, with $Y\sim 0.32$\ or even larger,
in agreement with what recently proposed by Joo \& Lee (2013). These stars should be
traceable on the MS of the cluster, once adequate photometric data is available.

Finally, we found several fast rotators. They are concentrated in a narrow region of the HB,
with $8400<T_{\rm eff}<9400$~K. There is strong correlation between rotational velocity
and temperature within our Group 1, which might suggest that fast rotators (where surface
rotation is assumed to be a proxy for core rotation, that is the parameter that might be
linked to position on the HB) are less massive than slow rotators, as proposed many years ago 
by Peterson et al. (1983). However, first, the difference in mass is very small 
($\leq 0.015$~M$_\odot$), so that core rotation is much less important than chemical
composition. Second, there is not any similar correlation between temperature and rotational
velocity for stars in the other groups. This seems to instead indicate that surface rotation 
can only be observed in a restricted range of temperatures along the HB of globular clusters. 
While the lack of significant rotation in cooler stars might be explained by their larger radius
and by magnetic braking, we have not an explanation for the slow rotation of the hotter
stars (see, however, Vink and Cassisi 2002 for a potential scenario). Further investigation
is required to establish that core rotation is indeed related to the colour of HB stars.

\begin{acknowledgements}
This publication makes use of data products from the Two Micron All Sky Survey, 
which is a joint project of the University of Massachusetts and the Infrared 
Processing and Analysis Center/California Institute of Technology, funded by the 
National Aeronautics and Space Administration and the National Science
Foundation. This research has made use of the NASA's Astrophysical Data System. 
This research has been funded by PRIN INAF ``Formation and Early Evolution of 
Massive Star Clusters".  We thank Philipp Richter for sending us the Str\"omgren
photometric data they obtained for M~22. VD is an ARC Super Science Fellow.
We thank an anonymous referee for suggestions that helped to improve the paper.

\end{acknowledgements}

\end{document}